\documentclass{aa}  
\usepackage{graphicx}
\usepackage{txfonts}
\usepackage{natbib}
\bibpunct{(}{)}{;}{a}{}{,}
\newcommand{\doceCO}{\mbox{$^{12}$CO}}
\newcommand{\doce}{\mbox{$^{12}$CO}}

\newcommand{\trece}{\mbox{$^{13}$CO}}
\newcommand{\treceCO}{\mbox{$^{13}$CO}}
\newcommand{\jsc}{\mbox{$J$=6$-$5}}

\newcommand{\jdu}{\mbox{$J$=2$-$1}}
\newcommand{\juc}{\mbox{$J$=1$-$0}}
\newcommand{\jdn}{\mbox{$J$=10$-$9}}
\newcommand{\jdq}{\mbox{$J$=16$-$15}}
\newcommand{\jdsq}{\mbox{$J$=16$-$15}}

\newcommand{\kms}{\mbox{km\,s$^{-1}$}}

\newcommand{\ms}{\mbox{$M_{\mbox{\sun}}$}}

\newcommand{\lsim}{\raisebox{-.4ex}{$\stackrel{\sf <}{\scriptstyle\sf \sim}$}}
\newcommand{\gsim}{\raisebox{-.4ex}{$\stackrel{\sf >}{\scriptstyle\sf \sim}$}}

\begin{document}
   \title{Warm gas in the rotating disk of the Red
     Rectangle: accurate models of molecular line emission
}

\titlerunning{The Red Rectangle rotating disk}

   \author{
          V. Bujarrabal\inst{1} \and J.\ Alcolea\inst{2}
 }

   \offprints{V. Bujarrabal}

   \institute{             Observatorio Astron\'omico Nacional (OAN-IGN),
              Apartado 112, E-28803 Alcal\'a de Henares, Spain\\
              \email{v.bujarrabal@oan.es}
\and
             Observatorio Astron\'omico Nacional (OAN-IGN),
             C/ Alfonso XII, 3, E-28014 Madrid, Spain
           }

   \date{Submitted: 28 November 2012; accepted: 27 February 2013}

  \abstract 
{} 
  {We aim to study the excitation conditions of the molecular gas in
    the rotating disk of the Red Rectangle, the only
    post-Asymptotic-Giant-Branch object in which the existence of an
    equatorial rotating disk has been demonstrated. For this purpose,
    we developed a complex numerical code that accurately treats
    radiative transfer in 2-D, adapted to the study of molecular lines
    from rotating disks.}
{We present far-infrared Herschel/HIFI observations of the \doce\ and
  \trece\ \jsc, \jdn, and \jdsq\ transitions in the Red Rectangle.  We
  also present our code in detail and discuss the accuracy of its
  predictions, from comparison with well-tested codes. Theoretical line
  profiles are compared with the empirical data to deduce the
  physical conditions in the disk by means of model fitting.  }
{We conclude that our code is very efficient and produces reliable
  results. The comparison of the theoretical predictions with our
  observations reveals that the temperature of the Red Rectangle disk
  is typically $\sim$ 100--150 K, about twice as high as previously
  deduced from mm-wave observations of lower-$J$ lines.  We discuss the
  relevance of these new temperature estimates for understanding the
  thermodynamics and dynamics of this prototype object, as well as for
  interpreting observations of other rarely studied post-AGB
  disks. Despite our sophisticated treatment of the line formation, our
  model cannot explain the relatively strong line-wing emission for
  intermediate-$J$ transitions. We argue that a model including a
  rotating disk only cannot reproduce these data and suggest that
  there is an additional extended (probably bipolar) structure
  expanding at about 7--15 \kms.  }
{}
\maketitle

\section{Introduction}

Disks orbiting post-Asymptotic-Giant-Branch (post-AGB) stars are often
postulated to explain the very energetic axial outflows that take place
in this evolutionary phase (see e.g.\ Bujarrabal et al.\ 2001; Soker
2002; Balick \& Frank 2002; Frank \& Blackman 2004). Planetary and
protoplanetary nebulae (PNe, PPNe) very often show axisymmetric shapes
and fast axial movements, which are thought to be caused by shock
interaction between the very collimated post-AGB jets and the slow and
isotropic AGB wind.  According to simulations, mass accretion from
disks rotating around post-AGB stars (with a magnetic field) can
provide the energy and momentum required to explain the PPN dynamics
via the ejection of very fast jets.  The study of these rotating disks
and their main physical conditions is therefore a basic requirement for
understanding the post-AGB ejections, and therefore the spectacular and
very fast evolution of the shape and dynamics of young PNe.

Disks/tori of molecular gas around post-AGB stars are commonly detected
as the central part of protoplanetary and young planetary nebulae,
although they are not observed to rotate, but to systematically 
expand (like the rest of the nebula). These expanding structures are
usually thought to be mere remnants of the former AGB winds.  See the
cases of M\,1--92 (Alcolea et al.\ 2007), M\,2--9 (Castro-Carrizo et
al.\ 2012), M\,2--56 (Castro-Carrizo et al.\ 2002), etc. Yet there is
one remarkable exception: the Red Rectangle. 

The Red Rectangle is a well-known PPN that surrounds the A1 star
HD\,44179, a spectroscopic binary (e.g.\ Waelkens et al. 1996). In the
visible and near infrared (NIR), the reflection nebula is very extended
(about 1$'$) and presents a conspicuous symmetry axis. It shows a
spectacular X-shaped morphology, including a remarkable series of
``ladder rungs'' perpendicular to the symmetry axis (Cohen et
al.\ 2004). Several models have been developed to explain these images,
attempting to derive the dust distribution and illumination pattern
(e.g.\ Men'shchikov et al.\ 2002; Koning et al.\ 2011) as well as the
dynamics of the jet-shell interaction that probably shaped the nebula
(Vel\'azquez et al.\ 2011 and references therein).

The Red Rectangle is the only PPN in which a central disk has been
actually observed to be in rotation (Bujarrabal et al.\ 2005), thanks
to high-resolution maps of the CO $J$=2--1 and $J$=1--0 mm-wave
lines. These maps show that the molecular gas in this source forms an
equatorial rotating disk perpendicular to the optical nebula
axis. Modeling of these observations indicates that the inner regions of
the molecular disk, closer than $R_{\rm kep}$ $\sim$ 8.4 10$^{15}$ cm,
are in Keplerian rotation around a central mass of about 1.7 \ms\ (for
a distance of 710 pc; see Bujarrabal et al.\ 2005).  In outer regions
of the disk, farther than $R_{\rm kep}$ and up to the outer disk radius
$R_{\rm out}$ $\sim$ 2.7 10$^{16}$ cm, the gas shows a slow expansion
of $\sim$ 0.8 \kms\ superimposed to rotation.

The orbiting material in the Red Rectangle was first proposed to
explain several properties of this object, namely the probable presence
of big grains in the nebula, the strong NIR excess, and the anomalous
abundances found in the stellar atmosphere, but without any direct
detection (see e.g.\ Waters et al.\ 1992; Waelkens et al.\ 1996; Jura
et al.\ 1997).  Following similar arguments, very compact rotating
disks have also been proposed to surround other evolved stars, mainly
in the post-AGB phase (see Jura \& Kahane 1999; Van Winckel 2003; De
Ruyter et al.\ 2006; etc.), but again no direct detection of the
velocity field has been obtained in them.  In one of these objects, 89
Her, mm-wave maps show a very compact component (smaller than $\sim$ 5
10$^{15}$ cm), whose low-velocity dispersion seems incompatible with
expansion and supports the existence of an inner Keplerian disk (see
Bujarrabal et al.\ 2007); this compact component is surrounded by an
extended hour-glass-shaped halo in slow expansion.

A relatively high temperature was deduced in the Red Rectangle disk
from the \jdu\ and \juc\ maps, with $T$ $\sim$ 65 K at $r$ = $R_{\rm
  kep}$ (8.4 10$^{15}$ cm, the distance at which the velocity regime
changes) and typical values of $T$ $\sim$ 100 K in the Keplerian inner
region. Note however that low-$J$ CO transitions are poor probes of the
kinetic temperature in such warm regions, because the \jdu\ line only
requires about 15 K to be excited (the \juc\ requiring 5.5
K). Therefore, although those observations are very useful for studying
the structure and dynamics of the warm molecule-rich regions, the
derived temperatures are just a crude approximation, and so the
excitation conditions in the Red Rectangle's rotating disk remained
poorly probed, especially at the innermost regions. This problem also
applies to the disk mass, though less critically, because errors in the
temperature estimate in general may induce errors in the mass estimate.
Very generally speaking, observations of transitions whose excitation
energy is comparable to or somewhat higher than the excitation of a gas
component, usually given by its kinetic temperature, are necessary to
properly study that component.

Herschel/HIFI observations of high-$J$ molecular lines in the Red
Rectangle have been recently published (Bujarrabal et al.\ 2012),
including data of the \doce\ and \trece\ \jsc, \jdn, and
\jdq\ transitions (all of them, excepted the \trece\ \jdq\ line, are
clearly detected).  The high spectral resolution of the heterodyne
spectrometer HIFI yields accurate line profiles from which we can
distinguish the contributions of different parts of the disk rotating
at different velocities.  The \doce\ $J$ = 6/10/16 rotational levels
are placed at 115/300/750 K above the ground level, and therefore can
properly probe the excitation of the warm gas in the disk. Bujarrabal
et al.\ (2012) concluded that the temperature deduced from the low-$J$
data was significantly underestimated, but using quite general
considerations on the line emissivity as a function of the temperature
and not a detailed modeling of the molecular excitation.

Transitions requiring still much higher excitation (over 3000 K) have
also been observed in the Red Rectangle using Herschel/PACS. Modeling
of such data is particularly uncertain and beyond the scope of this
paper. These lines must originate only from far inside and hot disk
regions that are not probed at all by the mm-wave maps (and therefore
are largely unknown), and the PACS observations do not provide
information on the profile shape.

Accurate modeling is particularly important for a rotating disk, whose
line excitation shows peculiar properties because it depends on photon
trapping in a velocity regime that yields radiative interactions
between very distant points and a quite complex structure of these {\em
  coherence regions}. Therefore, very complex calculations of radiative
transfer and level population are necessary to really approach the line
emission of such a structure. In this paper, we present an accurate
treatment of radiative transfer in one and two dimensions, applied to
the CO line emission from the Red Rectangle. The radiative transfer
calculations for each of the transitions, coupled with the statistical
equilibrium equations that give the populations of the relevant energy
levels, are solved by the code simultaneously for all points of the
molecule-rich cloud. Once the solution is attained after a complex
iterative process, the resulting level populations are used to
calculate the brightness distributions of a given line by solving the
standard radiative transfer equation. The characteristics of the code
are discussed in detail in Appendix A for both the 1-D and 2-D cases.
We present extensive tests to the code, particularly in comparison with
previous calculations and with results from the well-known
large-velocity-gradient ($LVG$) approximation, and discuss the validity
of different approximations to the line excitation under various
conditions. We conclude that our code yields accurate calculations in
the considered conditions, notably for a rotating disk.

We compare line profile predictions from our code with the
Herschel/HIFI data of the Red Rectangle, which were carefully
reanalyzed.  The physical conditions in the Red Rectangle disk are
deduced from model fitting, keeping the nebula model as similar as
possible to that deduced by Bujarrabal et al.\ (2005); we confirm that
the temperature was significantly underestimated from their analysis of
low-$J$ observations.

\section{Nonlocal treatment of radiative transfer and line excitation}

We have developed a completely new code to treat radiative transfer and
line excitation in one and two dimensions, adapted to interstellar
clouds and circumstellar shells or disks. It is well-known that
molecular line formation from a gas cloud is basically given by the
level populations in each point together with the cloud structure and
kinetics. The level population in a given point is given by the rates
of population from and depopulation to the other levels. We can assume
statistical equilibrium for each level (its population becomes
independent of time), because the typical rates are much faster than
the typical evolution times of the physical conditions in the
interstellar or circumstellar media.

The population transfer rates are mainly due to collisional and
radiative transitions. Collisional rates, $C$, are assumed to be
proportional to the local density (in our case, the gas is mostly
composed of H$_2$), since three-body encounters are very rare in our
case, and are assumed to satisfy the microreversibility principle,
which guarantees the Boltzmann distribution of populations for thermal
equilibrium (with the thermal bath, i.e., when only collisions are
relevant):
\begin{equation}
C_{l,u} ~=~ C_{u,l} ~g_u/g_l~ e^{-\frac{E_u-E_l}{kT}} ~~,
\end{equation}
where $l$ and $u$ represent the lower and upper levels of the
transition, $g$ and $E$ are the level statistical weights and energies,
and $T$ is the gas kinetic temperature. Other symbols have their usual
meanings.  Collisional rates in cm$^{3}$s$^{-1}$ units are usually
obtained from {\em ab initio} quantum mechanical calculations of
molecule encounters.

The transition rates of each radiative transition are given by the
Einstein $A$- and $B$-coefficients and the incident radiation at the
resonant frequency. In this code version for CO, we only include
allowed rotational transitions in the vibrational $v$=0 state that
satisfy $\Delta J$ = $\pm$ 1. The contribution of vibrational
  excitation to the emission of the $v$=0 rotational ladder is
  negligible for the high densities in the disk we are dealing with,
  see Sects.\ 2.2 and 4, though it may be relevant in other cases. Of
course, the treatment of the ro-vibrational lines is identical to that
of the rotational ones, and the code can take them into account if
necessary. The radiative transition probabilities are given by
\begin{equation} R_{u,l} ~=~ A_{u,l} ~+~ 4\pi~B_{u,l}~\overline{J}
\end{equation}
and 
\begin{equation}
R_{l,u} ~=~ 4\pi~B_{l,u}~\overline{J} ~~. 
\end{equation}
$\overline{J}$ is the intensity $I(\nu)$ that arrives at the point
integrated over all angles and frequencies (weighted with the
normalized line profile $\Phi$):
\begin{equation}
\overline{J} ~=~ \int J(\nu) ~\Phi(\nu) ~{\rm d}\nu ~~,~~ 
J(\nu) ~=~ \frac{1}{4\pi} \int I(\theta,\phi) ~{\rm d}\Omega ~~.
\end{equation}
The Einstein coefficients also satisfy the conditions necessary to
yield a Boltzmann distribution when equilibrium with a radiation thermal
bath takes place:  
\begin{equation}
B_{u,l} ~=~ \frac{c^2}{8\pi~h\nu^3}~ A_{u,l} ~~,~~ B_{l,u} ~=~
\frac{c^2}{8\pi~h\nu^3}~ \frac{g_u}{g_l}~ A_{u,l} ~~.
\end{equation}
The relations between the $A$ and $B$ coefficients may vary
from author to author, depending on the formulation taken for the rates
$R$, but the dependence of $R$ on $A$ must remain unchanged.

The (local) line profile $\Phi$ is given by the local velocity
dispersion and the corresponding Doppler shifts, since the natural line
width is completely negligible in our case in which collisional and
radiative transition rates do not exceed $\sim$ 1 s$^{-1}$.  We are
implicitly assuming complete redistribution, i.e., that the line
profile $\Phi$ is the same for absorption and emission; this
approximation simplifies the treatment and is again fully satisfied in
our case with very probable elastic collisions (line scattering is not
very important in the low-probability CO transitions we are
discussing).

The key problem is of course the calculation of $\overline{J}$ in each
point of the cloud, which must be performed by solving the usual
radiative transfer equation for resonant transitions, and therefore
depends on the conditions of the other points. This introduces a very
complex coupling between the excitation of the different parts of the
cloud, and accordingly the solution of the whole problem is difficult in the
general case. Several ways to solve it have been proposed, including
approximations of the integrals by means of quadrature formulae, Monte
Carlo methods, or approximate integration of the intensity obtained from
ray tracing (see for instance Lucas 1976; Bernes 1979; van Zadelhoff et
al.\ 2002). We adopt a ray tracing procedure, see Sect.\ 2.2.

Once $\overline{J}$ is calculated for all transitions, the code must
solve the statistical equilibrium equations for the population $n_i$
of all levels $i$ in all points of the cloud:
\begin{equation}
\dot{n}_i = \sum_{j} [n_jC_{j,i} + n_jR_{j,i} - n_iC_{i,j} -
  n_iR_{i,j}] = 0 ~.~
\end{equation}
The solution is a set of level populations for all points consistent
with the integrated intensity and transition rates. In practical cases,
we must use the level populations to calculate again $\overline{J}$ for
all transitions and points, and the system of equations results in an
iterative process, where  eqs.\ 4 and 6 are  alternatively solved.

Once convergence has been attained and the excitation of the treated
molecule has been calculated, one can calculate the resulting line
profiles by just solving the radiative transfer equation along the
lines of sight for a number of frequencies around the resonant one,
i.e., for a number of relative Doppler-equivalent velocities.

\subsection{The LVG or Sobolev approximation}

A simplified way to address the problem is possible when there is a
significant velocity gradient in the molecular cloud. Under certain
conditions (sufficiently high velocity and sufficiently high velocity
gradient and, for instance, monotonically increasing or decreasing
radial velocity), points separated by a significant distance do not
interact radiatively because the relative Doppler shift is much higher
than the local velocity dispersion. When the {\em coherence region}
(where radiative interaction actually takes place) around a given point
is small enough compared with the variation scale of the physical
conditions, only the local properties are relevant for calculating the
averaged intensity, which becomes
\begin{equation}
\overline{J} ~\sim~ (1 ~-~ \beta)~S ~+~ \beta_c I_c ~~.
\end{equation}
$\beta$ is the {\em escape probability}, which is a local function and
depends on the velocity field. In the case of a radial field, $\beta$
depends on the local logarithmic gradient of the velocity $V(r)$ ($r$
being the distance to the center): $\epsilon$ = $\frac{{\rm
    d~ln}(V)}{{\rm d~ln}(r)}$. In the simplest case, for $\epsilon$ =
1, the ({\em characteristic LVG}) opacity $\tau$ is isotropic and
\begin{equation}
\beta ~=~ (1-e^{-\tau})/\tau ~~.
\end{equation}
$\beta$ is in fact the probability that a photon emitted from a point
finally leaves the cloud in spite of (local) trapping; $\beta$ $\sim$ 1
in optically thin cases and $\beta$ $\sim$ 1/$\tau$ in the optically
thick limit.  $\beta_c~I_c$ represents the contribution of the
radiation from continuum sources, which can often be written in two
separated terms, the intensity of the continuum and $\beta_c$, a
generalized continuum escape probability that depends only on local
properties.

This approximation was first developed by V.\ Sobolev (Sobolev,
1960). A detailed and very understandable formalism can be found in
Castor (1970), and some generalizations were discussed by Bujarrabal et
al.\ (1980).

The {\em LVG} approximation enormously simplifies the problem, since
the population of the energy levels at a given point only depends on
local properties, though the system of equations (6) is strongly
non-linear and an iterative numerical process is necessary (except for
very low optical depths and obviously thermalized level systems). The
$LVG$ formalism is known to yield accurate solutions of the excitation
state, even when the conditions are barely satisfied, see discussion on
its validity in Appendix A. It contains most of the {\em ingredients}
relevant in the line excitation, including trapping effects, yields
very fast calculations, and, finally, does not require one to assume
details of the cloud structure, dynamics, and physical conditions, of
which we only know characteristic values in actual sources. In any
case, the $LVG$ codes can often be applied to estimate the molecular
excitation (i.e., the level populations), but they are not accurate
enough to estimate the resulting line intensity and profile, which must
be calculated by solving the exact transfer equation across the cloud
using the set of $n_i$ given by the $LVG$ code.  We compare $LVG$
results with our non-local calculations in several cases and discuss
the validity of this approximation in Appendix A.

\subsection{Our code}

As we show in Appendix A, for some velocity fields that are present in
situations of astrophysical interest, the molecular excitation cannot
be approached by means of simplified treatments, such as the $LVG$
formalism. This is the case of rotating disks, in which radiative
interaction between distant points takes place. We also show that some
excitation effects, such as the underexcitation often found at the
cloud edge, are similarly difficult to describe with these
approximations.  We have developed a new code to accurately solve the
full radiative transfer equations in one and two dimensions, coupled
with calculations of the level populations, whose general formalism has
been described above. The leading idea of our code is that the
geometrical parameters must be calculated as much as possible at the
beginning and then used in all iterations.

The cloud is divided into a high number of small cells, within which we
assume constant excitation conditions. The representative values of
$\overline{J}$ for all transitions and of the level populations $n$ for
all levels are calculated for the center of the cells. As mentioned, in
our code $\overline{J}$ is calculated following the ray-tracing
procedure. A sufficiently high number of ray directions are considered
that sample the whole space of the directions. We calculate the paths
of the model rays that traverse other cells (i.e., the lengths
traversed by the ray in each cell, the relative velocity between the
cells in the ray direction, etc.). The rays are defined by means of
their azimuth and elevation angles with respect to the symmetry axis of
the model nebula (our 2-D calculations assume axial symmetry).  This
procedure is not optimized for 1-D calculations in a spherical cloud,
but the method was kept since our goal is in fact its extension to 2-D
calculations. For an initial set of level populations, we calculate the
intensity $I$ for all rays arriving at each point and for a
sufficiently high number of frequencies within the local line profile,
solving the radiative transfer equation discretized in the high number
of cells traversed by the rays.  We then calculate $\overline{J}$ for
the considered point, approaching the integral in eq.\ 4 by the
addition (with the corresponding weights) of the intensities arriving
along the rays for all the considered directions and frequencies.  With
the set of values of $\overline{J}$ for all radiative transitions, we
calculate the level populations \{$n_i$\} (from eq.\ 6) and, using
them, a new iteration is performed. The efficiency of the iterative
process basically depends on the total opacity; typically, the number
of iterations is of the same order as the highest opacity in the cloud,
since the information is transferred from one point to another by steps
of about 1/$\tau$ of the total size in each iteration. The velocity
field is also important in the computing time. When the macroscopic
velocity is very high, many velocity points must be considered and, in
general, the {\em coherence region} (as defined in Sect.\ 2.1) becomes
very small and a very fine cell net is required; in this limit, an
$LVG$-like approach is obviously preferable.  We show details of
the computer code in Appendix A, where we also extensively compare the
results of our code with previous calculations and with $LVG$
calculations, with very satisfactory results.

Several conclusions can be reached from our detailed analysis in
Appendix A:

\noindent
{\bf 1.} Our code is fully consistent with previously published
calculations and with results obtained from the $LVG$ approximation,
when it can be applied. We conclude that the results from the code are
reliable.

\noindent
{\bf 2.} Accurate non-local calculations are necessary in the case of
rotating disks, in which the CO line formation cannot be adequately 
approximated with $LVG$ or $LVG$-like codes. 

\noindent
{\bf 3.} Nevertheless, $LVG$ estimates of the level populations (not of
the line profile) are reasonably accurate in many other cases, even
when the high-velocity gradient requirement is barely satisfied, and
can be used in a wide variety of situations.

Another conclusion can be reached from direct inspection of the
radiative transition probabilities and is confirmed by our
calculations: 

\noindent
{\bf 4.} The CO low-$J$ transitions are easily thermalized. Their low
Einstein A-coefficients, $\sim$ 10$^{-7}$ s$^{-1}$, guarantee
thermalization for densities $n$ \gsim\ 10$^{4}$ cm$^{-3}$
(e.g.\ Bujarrabal et al.\ 1997). Accordingly, our calculations confirm
that the previous analysis of the mm-wave maps by Bujarrabal et
al.\ (2005; where $LTE$ was assumed) are basically correct. However,
the Einstein coefficient increases very fast with $J$, approximately as
$J^3$; for instance, $A$(\jdq) = 4 10$^{-4}$ s$^{-1}$, and this line is
thermalized only for $n$ \gsim\ 5 10$^{7}$ cm$^{-3}$. Typical densities
in the object we study here, the Red Rectangle disk, are $n(R_{\rm
  kep})$ $\sim$ 4 10$^5$ cm$^{-3}$ (Bujarrabal et
al.\ 2005). Therefore, in our case (and in many others), thermalization
cannot be assumed for the high-$J$ lines and complex calculations are
necessary.

\section{Herschel/HIFI and IRAM 30m data of the Red Rectangle}

We present data of mm- and submm-wave lines of \doceCO\ and
\treceCO\ in the Red Rectangle, which are compared with model
predictions.  The data have been taken with the IRAM 30m telescope and
the HIFI instrument onboard the Herschel Space Observatory (HSO).  The
spectra are shown in Fig.\,1 (also Fig.\,3 and Figs.\,B1--B3).

Data for the \jdu\ line were taken with the EMIR receivers of the IRAM
30m telescope. These data are new and belong to a larger survey of
molecular gas disks around post-AGB objects, which will be published in
a forthcoming paper. The mm-wave data were calibrated in the
standard way at the 30m instrument in units of main-beam
Rayleigh-Jeans-equivalent temperature $T_{\rm mb}$. These values were
re-scaled after the observation of strong CO lines of well-known
emitters (CW\,Leo, CRL\,618, and NGC\,7027).  After these corrections,
we estimate the absolute incertitude in fluxes to be approximately
$\pm$20\%. More details on the observation will be given in the
paper dedicated to the disk survey.

Data for the \jsc\ lines and above were obtained with the Wide
Band Spectrometers of the Herschel/HIFI instrument. These data were
already discussed by Bujarrabal et al.\ (2012), where all details on
the observations and data reduction can be found. Here we present data
obtained after an improved reduction, which include the
most recent values for the calibration parameters of the instrument
(Roelfsema et al.\ 2012). Intensities are also given in units of
$T_{\rm mb}$. For the HIFI data, uncertainties in the calibration are
larger and mostly due to uncertainties in the side-band gain
calibration and the baseline, with values of about 20\% for the
\jsc\ and \jdn\ lines and of 30\% for \doceCO\ \jdsq.

A summary of the main line parameters (including the observed line
  peak, area, and uncertainties) is given in Table 1, together with
  some parameters of the model fitting.

\section{Physical conditions in the molecular disk of the Red Rectangle}

   \begin{figure}
   \centering \rotatebox{0}{\resizebox{9cm}{!}{ 
\includegraphics{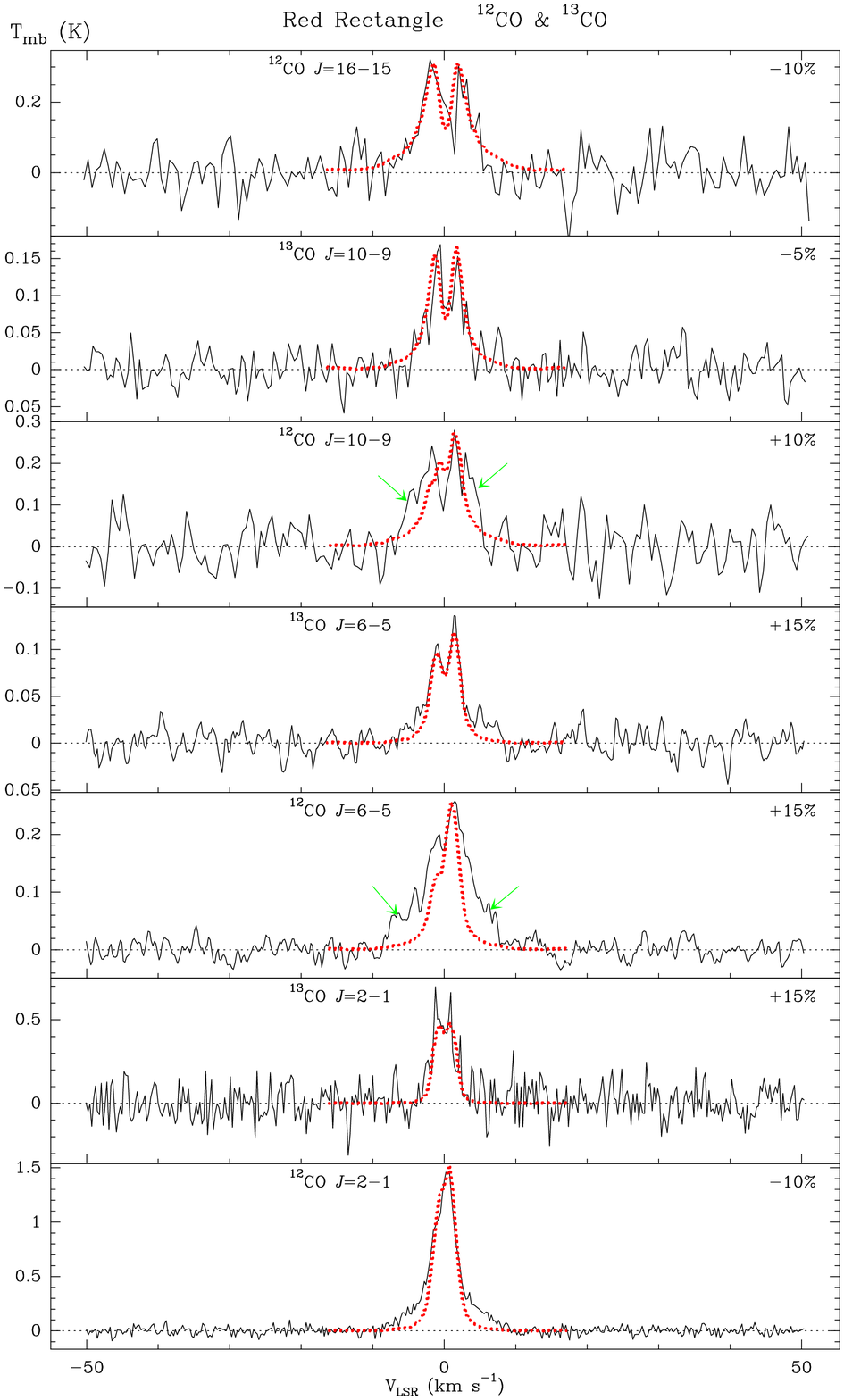}
}}
   \caption{Observed profiles of the  mm- and submm-wave transitions
     (black), Sect.\ 2. We also show (red points) 
     our best-fit model predictions from a
     rotating disk, which adequately reproduces the line cores,
     but underestimates the line-wing emission of some lines
     (green arrows), see Sect. 4. The values of the free scale 
     parameter in
     the model predictions that account for uncertainties in the
     calibration are indicated in the upper-right corners.  }
              \label{}%
    \end{figure}

\begin{table*}
\caption{CO lines: summary of observational parameters and fitting results}             
\label{tabcolines}      
\begin{tabular}{rlcccccccccc} \hline\hline
species  & line       & rest freq. & peak & area$\dagger$        
                                                       & rms$\ddagger$    
                                                                &resol. & HPWB       & $T_{\rm mb}$/$T^*_{\rm a}$
                                                                                            & Calib. & \multicolumn{2}{c}{Model} \\
         &            & (GHz)       & (K) & (K\,\kms) & (K)   &
(\kms) & ($''$)      &      & uncer. & scale factor & peak (K) \\ \hline
        
\doceCO  & $J$=16--15 & 1841.346    & 0.31  & 2.05      & 0.055     &
0.57   & 11\farcs6 & 1.4  &  30\%  & 0.90   & 0.31 \\ 
\treceCO & $J$=10--9  & 1101.350    & 0.17 & 0.81       & 0.02      & 0.41
& 19\farcs1 & 1.3  &  30\%  & 0.95   & 0.17 \\
\doceCO  & $J$=10--9  & 1151.985    & 0.27 & 1.77       & 0.05    & 0.52
& 18\farcs3 & 1.5  &  15\%  & 1.10   & 0.27 \\
\treceCO & $J$=6--5   &  661.067    & 0.14 & 0.66       & 0.012  & 0.23
& 32\farcs0 & 1.3  &  15\%  & 1.15   & 0.13 \\
\doceCO  & $J$=6--5   &  691.473    & 0.26  & 1.92       & 0.015     & 0.21
& 30\farcs8 & 1.3  &  15\%  & 1.15   & 0.26 \\
\treceCO & $J$=2--1   &  220.399    & 0.65  & 2.33      & 0.09     & 0.27
& 12\farcs5 & 1.7  &  20\%  & 1.15   & 0.5 \\
\doceCO  & $J$=2--1   &  230.538    & 1.47 & 7.07       & 0.035     & 0.25
& 12\farcs0 & 1.7  &  20\%  & 0.90   & 1.5 \\
\hline\hline
\multicolumn{11}{l}{$\dagger$ In the interval [--10:+10]\,\kms} \\
\multicolumn{11}{l}{$\ddagger$ For the given resolution  } \\
\end{tabular}
\end{table*}

   \begin{figure}
   \centering \rotatebox{0}{\resizebox{7cm}{!}{
\includegraphics{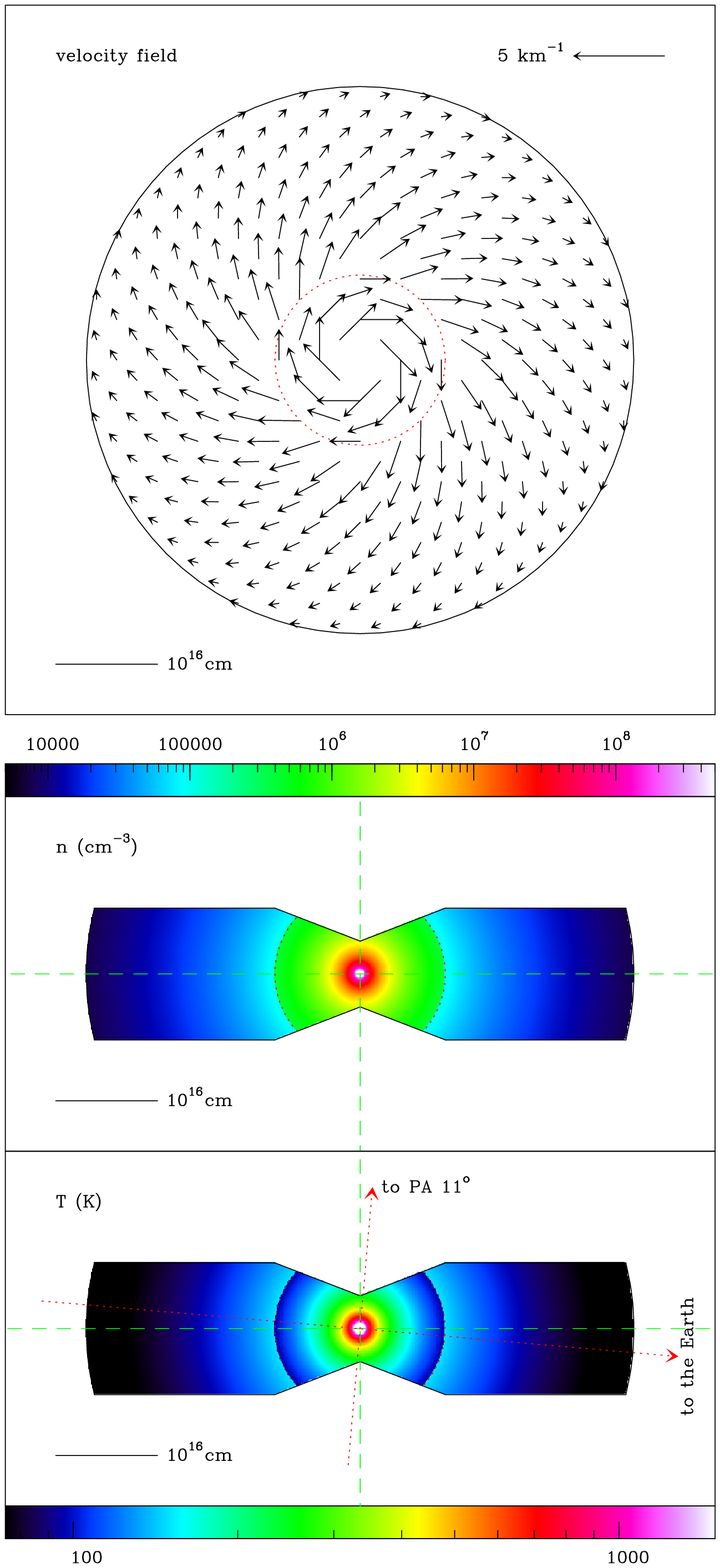}}}
   \caption{Velocity and physical conditions in the best-fit model
     disk. We represent the projection of the gas velocity on the
     equatorial plane (upper panel, disk seen face-on) and the density
     and temperature fields across the disk (seen edge-on). The red
     circle indicates the value of $R_{\rm kep}$, from which the
     dynamical regime changes (see text).}
              \label{}%
    \end{figure}

We have applied the most complex 2-D version of our code to fit our
observations of the \doce\ and \trece\ lines. A sophisticated treatment
is necessary because, as we have seen in Sect.\ 2.2, simplified
treatments ($LVG$ or $LTE$) are in-accurate.  In our numerical
calculations, the code typically considered more than 20 rotational
levels, about 800 cells, and about 200 rays (arriving at each cell); we
verified that convergence is attained for these values (as for the rest
of the numerical parameters entering the calculations). The
contribution of vibrational excitation to the emission of the $v$=0
rotational ladder is negligible in our case (as in most evolved
nebulae), even in the presence of continuum IR sources. A general
discussion on the effects of the vibrational excitation for PPNe and
PNe will be presented in Santander-Garc\'{\i}a et al.\ (2013, in
preparation), see a preliminary report in Santander-Garc\'{\i}a et
al.\ (2012). The excitation calculated from the code was used to
calculate the brightness of the observed transitions for a
high number of impact parameters in the direction of the line of sight
and of $LSR$ velocities. These distributions were convolved with
Gaussian beams to obtain predicted line profiles directly comparable
with the observations. Simulations of the interferometric observations
of \doce\ \juc\ and \jdu\ by Bujarrabal et al.\ (2005) were also
performed by convolution with the synthesized beams obtained in this
work (using an additional code similar to the one described in it); we
verified that for our new models the predicted low-$J$ maps are
compatible with these observations within the uncertainties.  In the
calculations presented here we used the most recent collisional rates
from the LAMBDA database (http://www.strw.leidenuniv.nl/$\sim$moldata,
Sch{\"o}ier et al.\ 2005; rates calculated by Yang et al.\ 2010) for
collisions with ortho- and para-H$_2$, for which we assumed an
abundance ratio of 3.  (Results are very slightly dependent on the
assumed ratio and we cannot be sure under which excitation conditions
H$_2$ was formed.)  Collisions with other gas components, notably He,
are not considered, so the derived density represents the total density
and not only the density of H$_2$.

\begin{table*}[bthp]
\caption{Structure and physical conditions in the molecular disk in the
  Red Rectangle derived from our model fitting of the CO data.
  Dependence on the assumed distance to the source is given in the
  relevant cases, but we favor a distance of 710 pc. We show in
  bold-face characters the values that change significantly compared to
  those deduced from the low-$J$ lines only; in practice this is only
  the gas temperature.}
\begin{center}                                          
\begin{tabular}{|l|cc|cc|}
\hline\hline
 & &  &  &  \\ 
 & \multicolumn{2}{c|}{Inner disk ($r < R_{\rm kep}$)} &  
\multicolumn{2}{c|}{Outer disk ($r > R_{\rm kep}$)}  \\ 
& & & & \\
{Parameter}  & {Law} & { Values} &   {Law} & {Values} \\ 
& & & & \\
\hline\hline
  &  &  &  &  \\
Outer radius  &   &  $R_{\rm kep}$ = 8.4 10$^{15}$ ($\frac{D({\rm pc})}{710}$)
 cm &  & $R_{\rm out}$ = 2.7 10$^{16}$ ($\frac{D({\rm pc})}{710}$) cm \\
  &  &  &  & \\
\hline
  &  &  &  &  \\
Disk & linear & {$H(R_{\rm kep})$ = 1.3 10$^{16}$
  ($\frac{D({\rm pc})}{710}$)} cm &
 constant & {$H$ = 1.3 10$^{16}$ ($\frac{D({\rm pc})}{710}$)} cm   \\
thickness & & {$H(0)$ = 6.5 10$^{15}$ ($\frac{D({\rm
      pc})}{710}$)} cm &   & \\
\hline
  &  &  &  & \\
Tangential & $V_t \propto 1/\sqrt{r}$ & ~~$V_t(R_{\rm kep})$ = 1.65 \kms
 & $V_t \propto 1/r$ & $V_t(R_{\rm kep})$ = 1.65 \kms  \\
velocity  & (Keplerian) & (central mass: 1.7 ($\frac{D({\rm
 pc})}{710}$) \ms) & (ang.\ mom.\ cons.) & \\
\hline
  &  &  &  & \\
Expansion & & 0 \kms\ & $V_{\rm exp} \propto \sqrt{a + b/r}$   
& $V_{\rm exp}(R_{\rm kep})$ = 1.6 \kms  \\
velocity &  & & & $V_{\rm exp}(R_{\rm out})$ = 0 \kms  \\
\hline
  &  &  &  & \\
Turbulence & constant & $\sigma_{\rm turb}$ = 0.2 \kms & constant &
 $\sigma_{\rm turb}$ = 0.2 \kms \\
velocity & & & & \\
\hline
  &  &  &  &  \\
 {\bf Temperature}  & $T \propto 1/r^{\alpha_T}$ & {\bfseries $T(R_{\rm kep})$ =
   95} K & $T \propto 1/r^{\alpha_T}$ & {\bfseries $T(R_{\rm kep})$ = 150} K  \\ 
 & & $\alpha_T$ = {\bf 1} & & $\alpha_T$ = 0.7 \\
  &  &  &  & \\
 Gas density & $n \propto 1/r^{\alpha_n}$ & $n(R_{\rm kep})$ = 4 10$^5$ 
($\frac{710}{D({\rm pc})}$) cm$^{-3}$ 
& $n \propto 1/r^{\alpha_n}$ & $n(R_{\rm kep})$ = 1 10$^5$
 ($\frac{710}{D({\rm pc})}$) cm$^{-3}$  \\
  &  & $\alpha_n$ = 2.2 &  &  $\alpha_n$ = 2.2  \\
\hline
\end{tabular}
\begin{tabular}{|l|cc|l|}
\hline
 & & & \\
{Other parameters}  & {Law} & {Values}
 & comments \\ 
& & & \\
\hline\hline
 & & & \\
Axis inclination from the plane of the sky & & 5$^\circ$ & from optical and
CO data  \\
 & & & \\
\hline
 & & & \\
Axis inclination in the plane of the sky (PA) & & 11$^\circ$ & from
optical and CO data \\
 & & & \\
\hline
& & & \\
Distance  & & between 380 and 710 pc & various arguments (Sect.\ 4) \\
 & & & \\
\hline
 & & & \\
\doce\ relative abundance & ~~constant~~ & ~~2 10$^{-4}$~~  & from CO data  \\
\trece\ relative abundance & ~~constant~~ & ~~2 10$^{-5}$~~  & from CO data  \\
 & & & \\
\hline
 & & & \\
LSR systemic velocity &  & $-$0.2 \kms & from CO data  \\
 & & & \\
\hline\hline
\end{tabular}
\end{center}
\end{table*}

As mentioned (Sect.\ 1; Bujarrabal et al.\ 2012), the emission from the
high-$J$ transitions in the Red Rectangle seems to be too intense to be
explained by the relatively low temperature deduced from our analysis
of low-$J$ lines (Bujarrabal et al.\ 2005), which are not suitable to
estimate the temperature in warm gas (Sect.\ 1). However, the model by
Bujarrabal et al.\ (2005) explained their high-resolution maps very
well, therefore we tried to keep the disk structure and dynamics
deduced in that paper as far as possible. Only the disk width was not
clearly determined from these data, because the deduced value was
smaller than the telescope resolution in the direction of the axis; the
(slight) inclination of the disk plane with respect to the line of
sight also hampers this estimation.

Calculations with our new code made with exactly the same nebula model
as deduced by Bujarrabal et al.\ (2005) confirm that too low emission
in the Herschel lines is then predicted, in particular the intensity of
the \jdq\ transition of \doce\ would be about 1/3 of the observed
intensity. It is therefore necessary to increase the temperature of the
central regions from which the high-$J$ transitions originate. The line
intensities are well reproduced if we increase the temperature by about
a factor two, to $T(R_{\rm kep})$ $\sim$ 100--150 K (we recall that
$R_{\rm kep}$ is a characteristic point in the disk at which the
dynamical regime changes from purely Keplerian to rotation plus a slow
expansion; see more details on the model and the parameter definition
in Bujarrabal et al.\ 2005). The physical conditions finally deduced
for the molecular disk in the Red Rectangle are shown in Table 2, in
which we indicate in bold-face characters the values that are
significantly different from those of the original nebula model by
Bujarrabal et al.\ (2005). Indeed, only the gas temperature field has
changed significantly compared to that model.  We had only to assume
one new parameter, the \doce/\trece\ abundance ratio, which was assumed
to be equal to 10. We compare in Fig.\ 1 the observed profiles and the
predictions using these parameters.  To account for observational
calibration incertitudes (Sect.\ 3), we introduced an intensity scale
tolerance in our fitting; the adopted scale factors are given in Table
1 (together with the expected calibration uncertainties) and Fig.\ 1
(in the upper-right corners of each panel).  The distributions of the
velocity, density, and temperature in the best-fit model nebula are
shown in Fig.\ 2.

Other mechanisms to increase the relative intensity of the \jdq\ line
are found not to be efficient enough to reproduce the observed
intensity values. For instance, high line opacities tend to increase
the population of the high-$J$ levels by means of photon trapping. The
\doce\ \jdq\ line is only moderately opaque in our case, but the
opacity of the intermediate-$J$ lines is high. This effect indirectly
leads to an increase of the $J$=16 and 15 level populations. Our
calculations, which accurately include the overexcitation effect due to
trapping, show that, under the expected conditions, this effect is not
enough by itself to explain the observed intensities. On the other
hand, we cannot increase the line opacity (for instance by assuming a
higher CO abundance) without affecting the intensities of the other
lines, and the calculations then contradict the observational data.

The model fitting, particularly in the line core, is satisfactory, see
Fig.\ 1; the differences between calculations and observations are
clearly smaller that the observational uncertainties. Because of the
many parameters involved in the very complex disk model (not all of
which are allowed to vary with the same degree of freedom) and the
various sources of uncertainty in the data (sometimes not described by
a normal distribution, such as the calibration uncertainty and the line
wings, which are not explained by our models), we did not try to find
the best fitting in a more systematic way, e.g.\ from a chi-square
approach.  There is, however, an obvious excess in the observed
profiles at {\em LSR} velocities $\sim$ $\pm$ 5--7 \kms\ (see arrows in
Fig.\ 1). We were unable to reproduce this line-wing excess, even
allowing variations of other disk parameters, see discussion in
Sect.\ 4.1.

In Appendix B we estimate the uncertainty of our temperature
determination, which is found to be about 20\% from a comparison of
predictions and observations for which we kept the other parameters
constant. We also show that these uncertainty values are not
significantly higher when we also allow moderate changes in the model
structure and in the gas density.  We recall that, in any case, the
temperature distribution and structure of the very inner region cannot
be well determined, because we severely lack spatial information on
them: the mm-wave maps by Bujarrabal et al.\ (2005) attained an angular
resolution equivalent to about 10$^{16}$ cm and were not very
sensitive to variations in the temperature, and, on the other hand, our
observations of FIR lines are useful for measuring the temperature but
have a still poorer resolution (though assuming Keplerian
rotation introduces a very useful but indirect relation between size
and velocity). We hope that future high-$J$, high-resolution mapping
will enable us to study the very inner regions of the Red Rectangle
disk better.

\subsection{Properties of the gas responsible for the line-wing
  emission excess}

As we have seen, it is possible to reproduce most of the features
identified in our profiles using a rotating disk model very similar to
that deduced from the low-$J$ observations, but with significantly
higher temperatures. However, our calculations indicate that the
relatively wide line-wings observed in $J$=6--5 (and less clearly in
other transitions, see green arrows in Fig.\ 1) are not explained by
models of this kind, even if we allow variations of parameters other
than the gas temperature or include new components. For instance, we
compare in Fig.\ 3 the observations and the model predictions using
the same physical conditions as for Fig.\ 1, but assuming an increase
by a factor 2 in the temperature and density in the disk region closer
than 2 10$^{15}$ cm to the axis. Evidently, the discrepancy in the
wings of the intermediate-excitation lines is somewhat lower but it
persists, and the intensity of the \jdsq\ line, mostly in its wings, is
now too high. Increasing the excitation of the inner disk regions even
more yields still much stronger wings in the \jdsq\ line.

   \begin{figure}
   \centering \rotatebox{0}{\resizebox{9cm}{!}{ 
\includegraphics{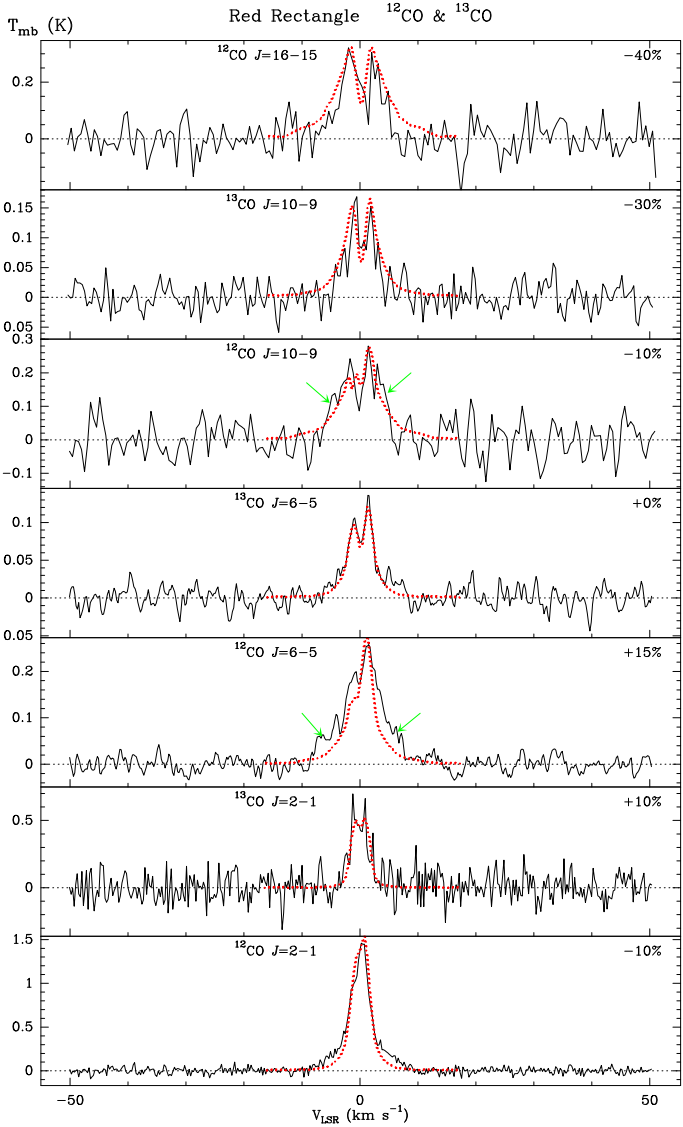}
}}
   \caption{Observed profiles of the mm- and submm-wave transitions
     (black) and predictions of our code (red points), assuming
     that the disk region closer than 2 10$^{15}$ cm to the axis is
     hotter and denser by a factor 2 than in our standard model, 
     Table 2. The symbols have the same meaning as in Fig.\ 1, note the
     values of the free scale parameter used in this case.}
              \label{}%
    \end{figure}

We think that a model that only includes a rotating disk of this kind
cannot explain the observed line-wing excess. A simple reasoning can
help to understand the reasons for this:

We have detected emission at a level of about $T_{\rm mb}$(\jsc) $\sim$
50 mK from gas rotating at velocities of up to about 7.5 \kms. If the
dynamics is the Keplerian one used here (or similar), such a velocity
only appears at a short distance from the star, $r$ $\sim$ 4 10$^{14}$
cm, which means an emitting region size of about 0\farcs07. Taking into
account the strong beam dilution at this frequency, the brightness
temperature in this region should be very high, $T_{\rm B}$ $\sim$ 9000
K. Except for very peculiar situations, the strongest line emission
from molecular gas appears when the level populations are thermalized
and the line is optically thick, then the brightness is equal to the
kinetic temperature. Therefore, the kinetic temperature in this region
must satisfy $T$ \gsim\ 10000 K to be able to account for the observed
intensity.

Such high temperatures are surprising for gas placed at 26 AU from this
A1-type star and one would not expect a high CO abundance in such a hot
gas, accordingly, this CO excitation state must be considered to be at
the limit of the possible situations. But even if these conditions are
really present, they would produce a too strong emission in the
\jdsq\ transition. The high opacity we must assume in the $J$=6--5
transition (from this small region) requires densities higher than
10$^8$ cm$^{-3}$ (in agreement with the density laws we are dealing
with), for which all transitions up to \jdsq\ must also be thermalized
and {\it a fortiori} be opaque, showing a brightness temperature
similar to that of \jsc. Taking into account the relatively low beam
dilution for the \jdsq\ observations (about seven times lower than for
\jsc), we expect $T_{\rm mb}$(\jdsq) $\sim$ 350 mK for $LSR$ velocities
$\sim$ $\pm$ 7 \kms. This evidently contradicts the observational data,
which show values lower than about 50 mK at these velocities and a line
peak of $\sim$ 300 mK.

Therefore, the only way to explain this wide emission in \jsc\ is to
assume that it comes from a more extended region, in which the
relatively high velocity cannot be dominated by rotation. The
excitation conditions can then be relaxed. We performed $LVG$
calculations for a very preliminary and simplified case. We assumed gas
at a typical distance from the center of 10$^{16}$ cm (therefore
occupying an angular region of about 3 arcsec$^2$), expanding at 15
\kms\ and with a temperature of 300 K; the density was assumed to be a
variable parameter.  Results are shown in Fig.\ 4. As expected, the
emission of this expanding gas component would be undetectable in
\doce\ \jdsq\ for moderate densities, not much higher than 10$^{5}$
cm$^{-3}$, at which \jdu, \jsc, and \jdn\ show an intensity of about
0.1 K. We took a typical velocity of about 15 \kms, assuming that the
observed $LSR$ velocities result from a high angle between the line of
sight and the velocity direction, but very similar results are
obtained, for instance, taking an expansion velocity of about 7
\kms. The calculated brightness values are very similar to the
brightness excess shown by the observed lines at $\pm$ 4--7 \kms\ $LSR$
and can then explain the observed features.  A still better agreement
can be found if we assume that the excess of the \jdn\ line at $\pm$ 4
\kms\ comes from a slightly denser region, while more diffuse gas would
also contribute to the excess in \jsc\ at $\pm$ 6 \kms; but a deeper
discussion of this very uncertain component is meaningless
and will only show these examples of the possible properties of the
emitting gas.

   \begin{figure}
   \centering \rotatebox{0}{\resizebox{9cm}{!}{ 
\includegraphics{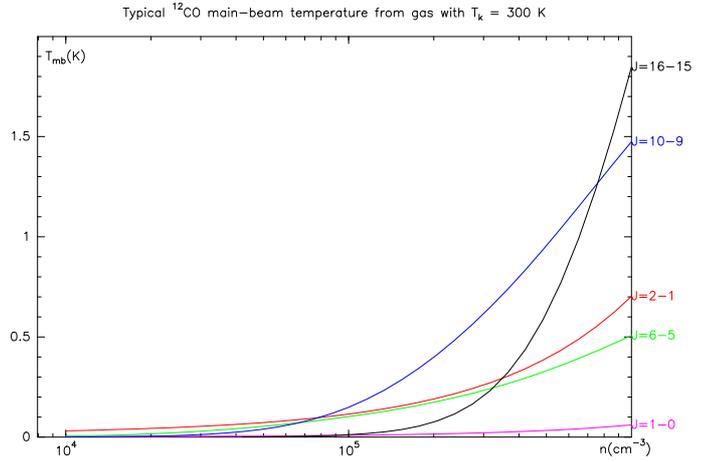}
}}
   \caption{Typical main-beam brightness temperature as a function of
     the density expected from the expanding gas component proposed to
     be responsible for the line-wing emission in
     intermediate-excitation CO lines (Sect.\ 4.1). }
              \label{}%
    \end{figure}

In a similar object, 89 Her, molecule-rich gas in expansion that forms
an hourglass-like structure has been mapped (Sect.\ 1, Bujarrabal et
al.\ 2007). In the Red Rectangle, the wide optical nebula is known to
be in expansion, although no CO emission was detected from it in the
interferometric observations by Bujarrabal et al.\ (2005), perhaps
because of the low expected brightness (\lsim\ 5 K for the example
presented above, which is lower than the continuum level) and its
relatively wide extent. A moderate increase in the sensitivity should
lead to the detection of this component.

In summary, we suggest that the emission excess found in the wings of
some observed lines, particularly obvious in our \jsc\ profile, is due
to the existence of a moderately extended component of molecule-rich
gas in expansion. No counterpart would be expected in very high-$J$
transitions. We do not attempt to describe this component in detail,
because we currently know too little about its properties.

\section{Conclusions}

We presented a new sophisticated code for calculating molecular line
excitation in two dimensions. The extensive tests presented in Appendix
A lead us to conclude that the numerical results are accurate. The code
in particular yields various line profiles (after convolving the
brightness distributions with the observational angular responses) that
can be directly compared with observational results.

We have used our code to study the physical conditions in the
equatorial disk of the Red Rectangle. This well-known protoplanetary
nebula is the only AGB or post-AGB nebula in which a disk in rotation
has been properly identified, thanks to mm-wave maps of
\doce\ \jdu\ and \juc\ by Bujarrabal et al.\ (2005). These authors
modeled the disk to reproduce the images and derived a reliable
description of its structure and dynamics. But such low-$J$
transitions, which require a very low temperature to be excited, are
not useful for studying the excitation state of the gas when its
temperature is \gsim\ 50 K, which is the case of most of the molecular
gas in the Red Rectangle.

Herschel/HIFI observations of \doce\ and \trece\ \jsc, \jdn, and
\jdq\ recently provided high S/N profiles of these transitions, see
Bujarrabal et al.\ (2012). These authors concluded from a preliminary
analysis of the excitation conditions that the kinetic temperatures
derived from the mm-wave maps were very probably underestimated. Here
we presented improved versions of these data.  We applied our new code
to disk models similar to those derived by Bujarrabal et al.\ (2005)
and compared the predicted profiles with the Herschel spectra (and
low-$J$ spectra and maps).

The characteristic temperatures derived by Bujarrabal et al.\ (2005)
were $T(R_{\rm kep})$ $\sim$ 65 K (decreasing with the radius as $T$
$\propto$ 1/$r^{0.7}$), where $R_{\rm kep}$ is the characteristic
radius at which the dynamics was found to vary from purely Keplerian to
rotation plus a slow expansion (the rotational velocity at $R_{\rm
  kep}$ is $\sim$ 1.65 \kms). We found that, as expected, the predicted
lines are too low for such temperatures, the \jdq\ line intensities
being about 1/3 of the observed ones. Assuming $T(R_{\rm kep})$ $\sim$
100--150 K (and keeping the rest of the parameters that describe the
structure and physical conditions in the disk practically unchanged),
we were able to reproduce the line cores, within a width of about 8
\kms, see Fig.\ 1. The resulting disk model is summarized in Table 2
(for details on the model and on meaning of the parameters, see
Bujarrabal et al.\ 2005). As we discussed in Sect.\ 4 and Appendix B,
the uncertainty in this determination of the typical temperature is
moderate, $\sim$ 20\%.

A proper measurement of the temperature in the rotating disk of the Red
Rectangle is of particular importance because the physical conditions
in these components, crucial for studying the post-AGB ejections and
evolution (Sect.\ 1), are still largely unknown and the physical
processes active in them, in particular the thermodynamics, are poorly
understood. Indeed, the Red Rectangle disk is practically the only one
that has been studied to date.  A good estimate of the gas
thermodynamics is also fundamental for understanding the complex
dynamics of post-AGB disks. A slow expansion was found in the outer
parts of the Red Rectangle disk from CO \jdu\ maps (Bujarrabal et
al.\ 2005), with velocities of about 1 \kms, comparable to the
rotational velocities and much lower than those typical of expanding
PNe and PPNe.  For temperatures \gsim\ 100 K, the thermal velocity
dispersion becomes \gsim\ 1 \kms; when the microscopic velocity is
comparable to the rotation velocity (and then to the escape velocity),
evaporation becomes able to affect the disk dynamics and introduces a
noticeable expansion (e.g.\ Gesicki et al.\ 2010, Hollenbach et
al.\ 2000) and could be the mechanism that drives the expansion
observed in the disk.  Finally, a good knowledge of the properties of
the Red Rectangle disk is also important to extend these studies to
other similar objects, since CO mm-wave (still unpublished)
observations are revealing a surprisingly high number of extended
rotating disks in certain types of post-AGB objects. The interpretation
of these data for sources for which we have significantly less
information than for the Red Rectangle requires a good knowledge of the
properties of the disk in this prototype. Estimates of the disk mass
and angular momentum from single-dish mm-wave lines, for instance,
cannot be performed for these nebulae if we have only a poor idea of
the characteristic temperature.

Even with these new temperature values, the comparison of our synthetic
profiles with those observed in the Red Rectangle is not fully
satisfactory. The line wing emission, coming from relatively fast gas,
is underestimated in intermediate-excitation lines, notably in the
emission of the \jsc\ transition at {\it LSR} velocities of about 5--7
\kms. We showed (Sect.\ 4.1) that this excess cannot be explained
assuming the Keplerian or Keplerian-like dynamics expected in the disk,
since then the emitting region must be very small and dense, yielding
also a very strong intensity in the \jdsq\ transition (we would expect
$T_{\rm mb}$ $\sim$ 350 mK at $\pm$ 7 \kms), which is in contrast to
the relatively weak wings of this line. This peculiar behavior of the
line-wing emission could be explained by postulating a
relatively extended and diffuse gas component (typical radius $r$
$\sim$ 10$^{16}$ cm, equivalent to $\sim$ 1$''$, and density $n$ $\sim$
10$^5$ cm$^{-3}$) that shows moderate typical velocities (between 7 and
15 \kms, depending on the projection angle) probably due to
expansion. This component, not to be confused with the slow disk
  expansion mentioned above, would be associated to the optical nebula
and be similar to the expanding structure found in the similar object 89
Her (Sect.\ 1, Bujarrabal et al.\ 2007). We discussed (Sect.\ 4.1) that
its mm-wave emission is probably too weak to be detected in our
previous mm-wave maps of the Red Rectangle (Bujarrabal et al.\ 2005),
but might be detected in future mapping with higher sensitivity.


\bigskip
\begin{acknowledgements}
We are grateful to Mario Tafalla for his help in understanding and
checking the Monte Carlo calculations. This work has been supported by
the Spanish MICINN, program CONSOLIDER INGENIO 2010, grant ``ASTROMOL"
(CSD2009-00038).  
\end{acknowledgements}

{}

\appendix

\section{Code tests and details}

Our code is designed to treat the excitation of rotational CO levels
and other simple molecules in a variety of conditions actually present
in the analysis of astronomical observations (Sect.\ 1). We therefore
focus our tests on the validity of the calculations in these cases. It
is well known that in quite extreme cases, such as very opaque
molecular transitions (particularly of HCO$^+$, van Zadelhoff et
al.\ 2002), the line formation is very difficult to treat and
predictions can significantly vary when different codes are used.  The
treatment of these cases is strongly dependent on the numerical
treatment of the line profile, the discretization in cells, etc. Even
worse, the results dramatically depend on the properties of the
discussed object (size of the cloud, variations of the physical
conditions across it, etc), which in actual cases are not well
known. In this context, when opacities are so high, the observational
parameters do not practically depend on the properties of the inner
regions, but just on the (very complex) structure of a very thin cloud
'photosphere', a surface layer that can hardly be adequately described.
We therefore focus on tests that request difficult convergence and
accurate calculations, but avoid extreme, hardly useful cases.

\subsection{1-D models}

We first compared the results from our code with previously published
calculations of well-tested codes in 1-D cases. For this purpose, we
used the 1-D version of the program. For comparison, we used
calculations published by Lucas (1976) and Bernes (1979), who used
well-known codes and applied them to CO emission in spherical
collapsing clouds. The Bernes code is a Monte Carlo approach that has
been extended to other molecules and a variety of physical conditions
and is very widely used today. The code by Lucas is based on a discrete
treatment of the frequencies and directions that approximates the
derivatives by finite (very small) differences and the integrals by
quadrature forms. To compare results from both papers with our
calculations, some {\em interpretation} of the assumptions and
approaches in these papers is necessary.

We first reproduce the calculations by Lucas (1976). We chose as an
example the CO $J$=2--1 and $J$=1--0 profiles calculated for his
low-density case and outer velocity equal to 1 \kms. The
local velocity dispersion (1-$\sigma$) is equal to 1 \kms. This is the
most representative case, since the level populations are not
thermalized and the profile shows a clear self-absorption due to the
excitation variation across the cloud. In this case the (constant)
physical conditions are density $n$ = 10$^3$ cm$^{-3}$, temperature $T$
= 30 K and CO relative abundance $X$(CO) = 10$^{-4}$. We suppose that
the published profile is given in units of Rayleigh-Jeans-equivalent
temperature (otherwise it is difficult to understand that the 2--1
line, mostly thermalized and opaque, shows this low peak temperature)
and that the background cosmic continuum was subtracted from the
published profile; both choices are common when predictions are to be
compared with radioastronomical observations.

We also assumed that the collisional rates used by Lucas (1976) are 
those published by Green \& Thaddeus (1976; the Lucas paper was
published when these calculations were still {\em in press}). We took
only the downward collisions from Green \& Thaddeus, the excitation
rates were calculated following the microreversibility principle
(Sect.\ 2), but we are not sure that the same was done by
Lucas. Finally, we calculated the collisions for the gas temperature,
30 K, by interpolating those published for 20 K and 40 K.

Following all these recipes, we resolved the excitation state of the CO
levels and calculated the resulting $J$=1--0 and $J$=2--1 profiles,
which are shown in Fig.\ A.1. We checked that convergence is reached in
the several parameters used in the code (number of cells, number of
considered rays, etc; see Sect.\ 2.2 and A.2). Clearly, the profiles
are identical to those given by Lucas (1976) up to the accuracy that
the comparison with the published profiles allows, despite the
uncertainties and possible sources of error that remain in the
comparison of both codes. We therefore consider the test satisfactory.

   \begin{figure}
   \centering \rotatebox{0}{\resizebox{9cm}{!}{ 
\includegraphics{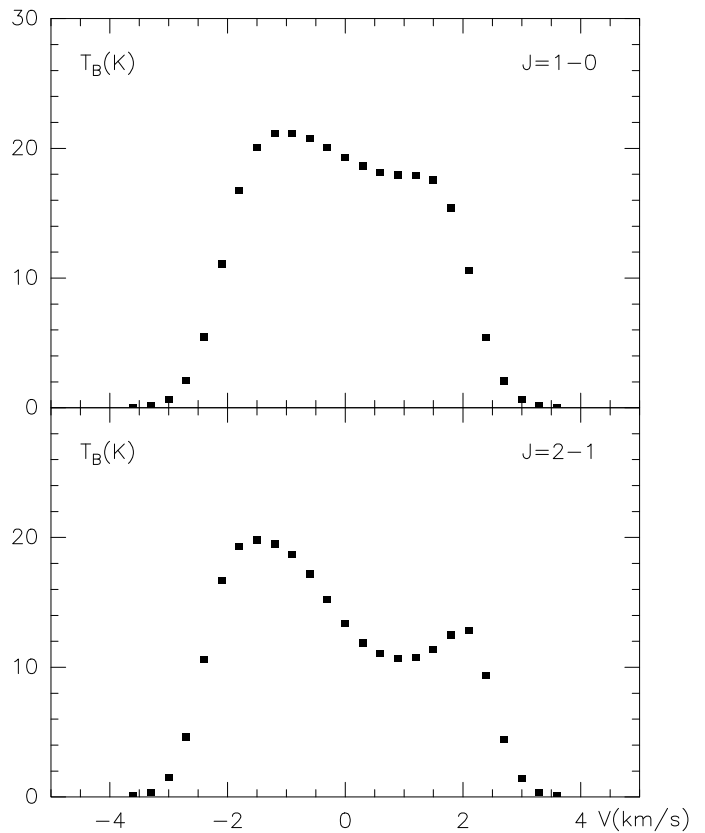}
}}
   \caption{Predicted profiles obtained with our code for the
     \doce\ \juc\ and \jdu\ lines, for the physical conditions used by
     Lucas 1976 (the case with density of 10$^3$ cm$^{-3}$ and a
     maximum radial velocity of 1 \kms), to be compared with
     profiles in Figs.\ 1 and 2 of Lucas (1976).}
              \label{}%
    \end{figure}

The comparison with calculations by Bernes is more complete because
this author also published the distribution of the excitation
temperature. The model nebula is somewhat larger (with a radius of 3
10$^{18}$ cm instead of 10$^{18}$ cm). Other parameters are $n$ = 2
10$^3$ cm$^{-3}$, $T$ = 20 K, and $X$(CO) = 5 10$^{-5}$. The velocity
field is the same as in Lucas (1976); we assume that the local velocity
broadening used by Bernes is also a 1-$\sigma$ dispersion. We also
applied the microreversibility principle in calculating the collisional
rates.

The predicted profile and excitation temperature ($T_{\rm ex}$) for the
CO $J$=2--1 transition derived with our code are shown in Figs.\ A.2
and A.3, to be compared with the corresponding figures in Bernes
(1979). We assume that Bernes (1979) gives the profile in units of
brightness temperature, in view of the zero-level (equal to about 2.7
K) and because the profile peak is equal to the kinetic temperature (as
expected in thermalized and opaque lines).  Clearly, our results are
identical to those obtained by Bernes, within the uncertainties,
including the numerical noise that is not negligible in the Monte Carlo
approach.

   \begin{figure}
   \centering \rotatebox{0}{\resizebox{9cm}{!}{ 
\includegraphics{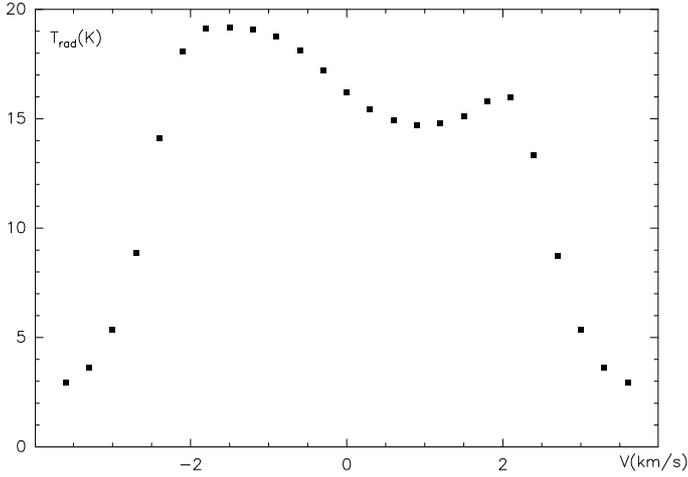}
}}
   \caption{Predicted profile obtained with our code for the
     \doce\ \jdu\ line, for the physical conditions used by
     Bernes (1979), to be compared with the corresponding profile in
     his Fig.\ 3.}
              \label{}%
    \end{figure}

   \begin{figure}
   \centering \rotatebox{0}{\resizebox{9cm}{!}{ 
\includegraphics{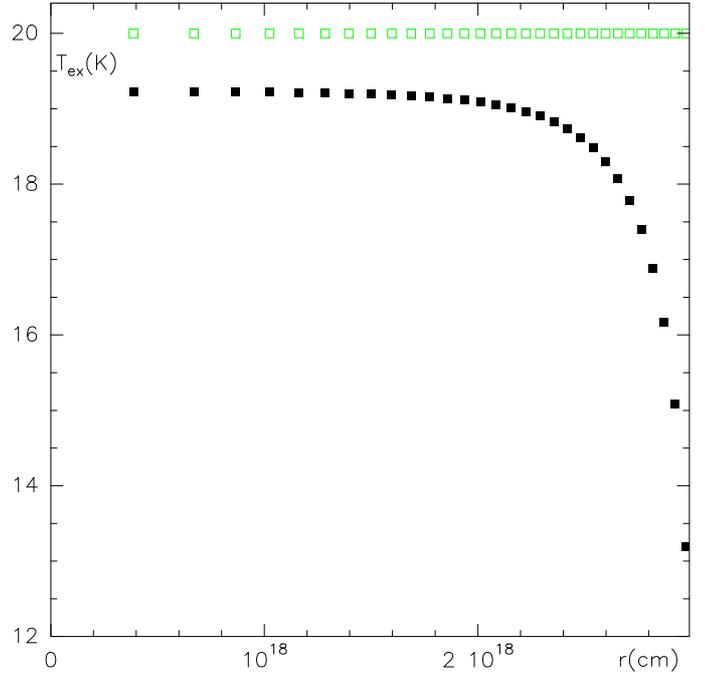}
}}
   \caption{Predicted excitation temperature obtained with our code for the
     \doce\ \jdu\ line for the physical conditions used by
     Bernes (1979), to be compared with his Fig.\ 2. The green, open
     squares represent the kinetic temperature of the cells.}
              \label{}%
    \end{figure}

We note that these two examples are very demanding tests for the
treatment of radiative transfer and its effects on the molecular level
excitation (the key problem), close to the too complex and scarcely
practical problems discussed before. The opacities are high, of about
50, and therefore the interaction between different points in the
nebula is very complex and the iterative process becomes slow. The
differences found between our calculations and those of the previous
codes are $\sim$ 1\% (often indistinguishable from the figures); the
largest difference, $\sim$ 3\%, was found in the comparison with the
excitation temperature of $J$=1--0 predicted by Bernes (1979) for the
outermost boundary of his model cloud (outer 10$^{17}$ cm, where
complex population cascades lead to some increase of the 1--0
excitation). We verified that differences found in various runs of our
code with different values of the numerical parameters are smaller than
1\%; we show in Sect.\ A.2 that in some cases our code may show a
numerical noise of about 1\% in excitation temperature, which could
also appear in the calculations shown here, although it does not
obviously manifest itself when different runs are compared. Most
probably the main differences with respect to previous works are due to
our interpretation of the previous calculations.

Differences of a few percent have often been found when very different
codes were compared, see Bernes (1979) and van Zadelhoff et
al.\ (2002). Their origin, except for the numerical noise in Monte
Carlo calculations, is difficult to know, as concluded by the quoted
authors.

The fact that the tests of our 1-D code are positive convincingly shows
that it treats the problem correctly. 

\subsection{Comparison between 1-D and 2-D codes}

We have mentioned (Sect.\ 2) that our code easily evolves to a 2-D
treatment, since, in some way, the two dimensions are included in the
code structure. We therefore built a 2-D code in which axial symmetry
is assumed and the cells are defined in terms of distance to the axis
of symmetry and to the equator. We assumed symmetry with respect to the
equator in the examples we show here, but it is not necessary in
the code (when this symmetry is assumed only one half the cells are
needed).

   \begin{figure}
   \centering \rotatebox{0}{\resizebox{9cm}{!}{ 
\includegraphics{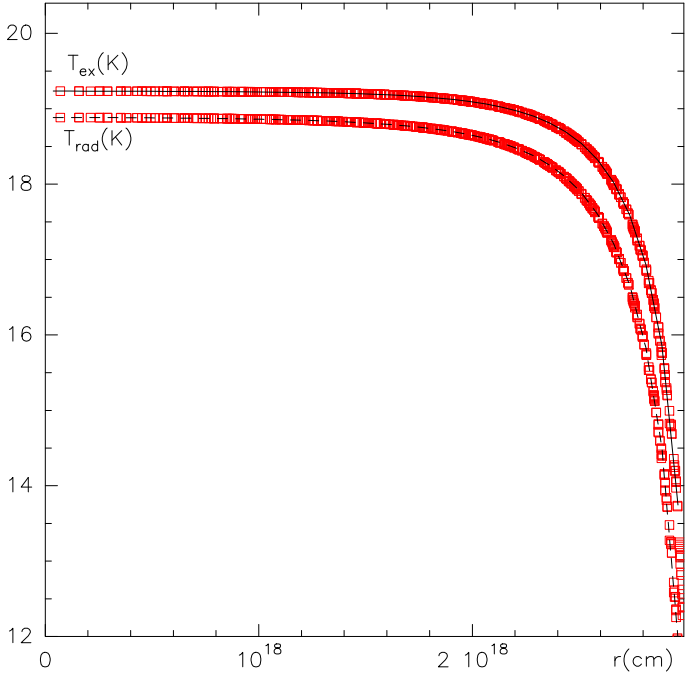}
}}
   \caption{Comparison between the excitation temperature ($T_{\rm
       ex}$) and the equivalent radiation temperature ($T_{\rm rad}$)
     of the \jdu\ transition calculated with the 1-D version of the
     code (continuous and dashed lines) and with the 2-D version (red,
     empty squares). Calculations performed for the same model as for
     Fig.\ A.3 (from Bernes 1979).}
              \label{}%
    \end{figure}

The tests of the 2-D code were performed running the 2-D code with the
same conditions as assumed in the 1-D examples. In other words, the
cell definition and radiative transfer treatment were performed fully
in 2 dimensions, with their coordinates given by the distances to the
{\em axis} and {\em equator}, but the physical conditions in the cloud
satisfy spherical symmetry. In this way, the 2-D treatment can be
checked by comparing it with 1-D code results, whose accuracy was
discussed above (Sect.\ A.1). A few non-local 2-D calculations of CO
excitation have been published for cases similar to those we are
interested in, e.g.\ Hogerheijde \& van der Tak (2000), but the lack of
details on the calculation and nebula model prevents a careful
comparison. These authors also checked their code by comparing its CO
predictions with the 1-D results from Bernes (1979).

In Fig.\ A.4 we compare the predictions of the
excitation temperature of the CO $J$=2--1 transition derived from our
1-D code (continuous black line) and those from the 2-D code (open red
squares). We used the same nebula model as in Sect.\ A.1 to compare our
results with those by Bernes (1979), including the collisional
rates. We considered a total of 40 cells in the 1-D calculations and
900 cells in the 2-D case.  We also show in this figure the equivalent
temperature of the radiation field ($T_{\rm rad}$) seen in each cell
(lower discontinuous, black line and red squares); of course, the
excitation temperature is placed between this radiation field
temperature and the kinetic one (20 K), its value is given by the
collisional/radiative de-excitation probability ratio.

The comparison is again very positive. We can see some numerical errors
(differences in excitation temperatures for cells at the same distance
to the center, between different cells of the 2-D code and with respect
to the 1-D calculations), but they are very moderate, \lsim\ 1\%. The
errors in $T_{\rm ex}$ are comparable to those found for $T_{\rm rad}$,
and calculating the radiation intensity in just one point per cell
is probably the source of the whole uncertainty.

   \begin{figure}
   \centering \rotatebox{0}{\resizebox{9cm}{!}{ 
\includegraphics{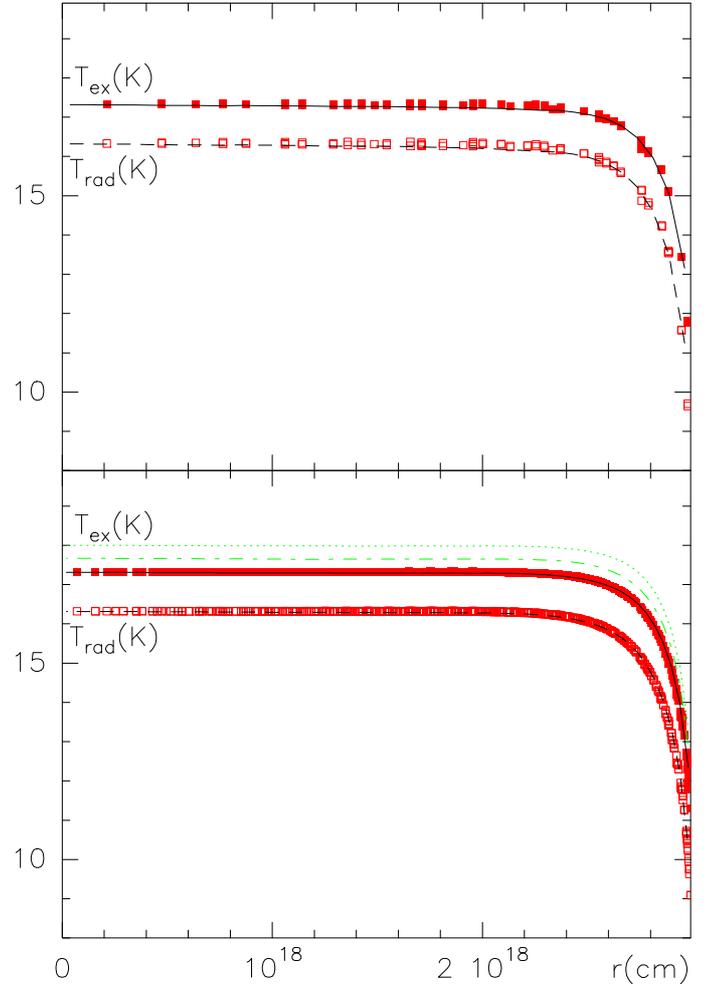}
}}
   \caption{Calculations similar to those performed for Fig.\ A.4, but
     for a model cloud in which the highest inward velocity is equal to
     4 \kms. Upper panel: using 40 and 100 cells for the
     1-D and 2-D calculations. Lower panel: using 75 and 900 cells; the
     green point and dash-point lines represent 1-D calculations when
     more recent collisional rates are used (see text). Note the
     improvement in the calculation noise when more cells are used and
     the moderate influence of collisional rates. }
              \label{}%
    \end{figure}

We also performed calculations for conditions under which the
errors may be slightly larger. When the macroscopic velocity field
becomes higher, the velocity variations within a given cell can be
relatively large. In such cases, errors may appear in the calculation
of the relative velocities of two cells, resulting in errors in the
calculations of the emission and absorption rates at the relevant
frequencies.

In Fig.\ A.5 we show the excitation and radiation temperatures for the
same case as before but increasing the maximum velocity to 4 \kms. For
calculations in the upper panel, we took 40 and 100 cells for the 1-D
and the 2-D models respectively (symbols are the same as in Fig.\ A.4).
As we see, differences slightly higher than 1\% are noticeable in the
2-D calculations and with respect to the 1-D results. Even some
systematic underexcitation of the 1-D calculations can be seen in the
figure. As mentioned, the reason for the appearance of some numerical
noise is that the cells are relatively large (which is particularly the
case of the spherical cells). The results converge for higher numbers
of cells, as we can see in Fig.\ A.5, lower panel, for which we used 75
and 900 cells for the 1-D and 2-D cases. In this panel we also show 1-D
calculations using more recent collisional rates (pointed, green line),
taken from the LAMBDA-database
(http://www.strw.leidenuniv.nl/$\sim$moldata; Wernli et al.\ 2006). As
we see, the results are different from those using calculations by
Green \& Thaddeus (1976), as in our previous runs, but not
dramatically, by about 5\% in $T_{\rm ex}$. From now on, we use these
more accurate rates in our tests. We finally show still more recent
calculations from the same database (Yang et al.\ 2010), dash-point
line; the differences are still smaller, \lsim\ 2\%, showing the
moderate changes in the final calculations often found when new
collisional rates are incorporated.

For higher values of the macroscopic velocity, convergence becomes
increasingly difficult. As we show below, in these cases calculating
the level populations with the well-known $LVG$ approximation is much
more efficient.

   \begin{figure}
   \centering \rotatebox{0}{\resizebox{9cm}{!}{ 
\includegraphics{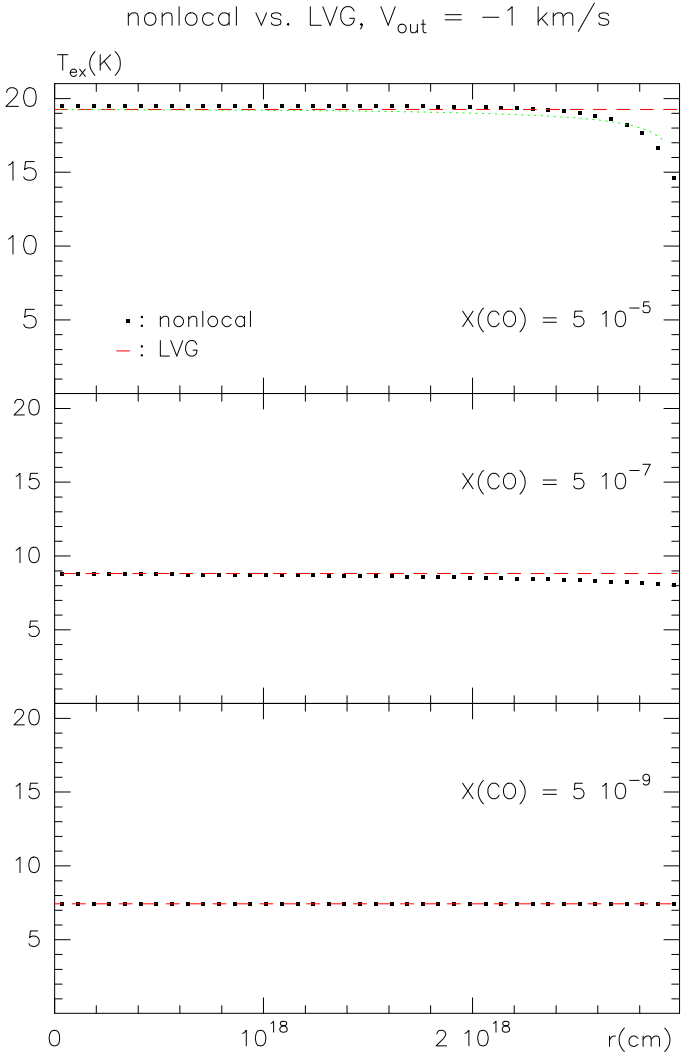}
}}
   \caption{Comparison between calculations performed using our
     non-local treatment of radiative transfer (squares) and the $LVG$
     approximation for a cloud model similar to that used previously
     (from Bernes 1979) for different values of the relative CO
     abundance (which led to an optically thin case for $X$(CO) = 5
     10$^{-9}$, lower panel). The green dotted lines represent a
     modification of the $LVG$ approach trying to simulate boundary
     effects.  }
              \label{}%
    \end{figure}

\subsection{Comparison with $LVG$ calculations}

The well-known $LVG$ approximation holds when in a cloud the module of
the macroscopic velocity increases monotonically with radius, with a
gradient steep enough to ensure that points decoupled radiatively are
still close and expected to show similar physical and excitation
conditions.  As mentioned in Sect.\ 2, the averaged radiation intensity
in any point, which rules the radiatively stimulated rates in the
transitions, is then a local parameter and the treatment becomes
equivalent to the definition of an escape probability of photons, which
represents the percentage of photons emitted by the gas that actually
escapes from the cloud in spite of self-absorption (trapping).  $LVG$
codes are therefore very fast and relatively simple, though they
incorporate the main ingredients of the problem.  It is also known that
this approximation yields reasonable values of the level population
even when the conditions are only marginally satisfied (although the
resulting line profile must in any case be calculated by solving the
exact radiative transfer equation for the derived level
populations). Because of the inherent radiative decoupling introduced
by the $LVG$ approximation, it can easily be extended to 2- or even 3-D
calculations, the (local) excitation being in fact independent of the
geometry of the cloud at large scale.

We performed $LVG$ calculations for the nebula models we studied in the
previous subsections and other similar ones. In the $LVG$
approximation, the effects of the nebula size and macroscopic velocity
field enter the calculation of the opacity via the factor $r/V$, $\tau$
$\propto$ $r/V$, where $r$ and $V$ are the distance to the center and
the expansion or collapse velocity at this distance; the local velocity
dispersion is neglected. Since we will apply this treatment to cases in
which the local velocity is not negligible compared to the macroscopic
one, we substituted in our calculations the value of $V$ by $V+\Delta
V/2$, where $\Delta V$ is the width at half-maximum of the local
velocity dispersion.

In a first case, we considered the same nebula model as Bernes (1979),
see Sect.\ A.2. We recall that this time we did not use the collisional
rates by Green \& Thaddeus but the more recent calculations
(Sect.\ A.3). In Fig.\ A.6, upper panel, we compare results from our
non-local 'exact' treatment (black squares) with those from the $LVG$
approximation (red dashed line). The $LVG$ results are surprisingly
similar to those from the non-local code, even if the highest
macroscopic velocity is just comparable to the local dispersion. In
reality, the $LVG$ predictions are independent of the point of the
nebula we are considering, because in this nebula model the value of
$r/V$ is constant. Therefore, in the outer regions, where very little
photon trapping occurs because we are very close to the nebula
boundary, the non-local code gives a somewhat lower excitation
(approaching the values obtained for lower abundances, see below) and
the $LVG$ limit of very low local velocity dispersion cannot deal with
this effect. It is possible to define 'equivalent' values of $r$ to
simulate this effect in the $LVG$ code and improve the predictions in
outer regions.  We do not discuss these procedures here, but just show
(pointed green lines) calculations in which we took for the outer
points an {\em equivalent radius} of the cloud equal to the distance
between the point and the outer circumference in the direction
perpendicular to the radius direction between the point and the cloud
center; we point out that the results are somewhat improved in the
outer regions but not yet fully satisfactory.

In the other panels of the figure, we show the same calculations for
lower values of the CO relative abundance, and therefore of the
opacities. As expected, the excitation temperature decreases (because
of the less frequent trapping) and the agreement with $LVG$ improves
(because in the optically thin limit the statistical equilibrium
equations, which give the level populations, become independent of the
radiative transfer treatment).

We have seen that when the local and macroscopic velocities are
practically the same, far from the $LVG$ assumptions of negligible
local $\Delta$$V$, the predictions of the simple $LVG$ codes are very
reasonable. Only the decreasing excitation toward the edge of the cloud
in very opaque cases can hardly be accounted for by an $LVG$ treatment,
and therefore the asymmetries of the profiles seen in Sect.\ A.1 cannot
be obtained from it.

   \begin{figure}
   \centering \rotatebox{0}{\resizebox{9cm}{!}{ 
\includegraphics{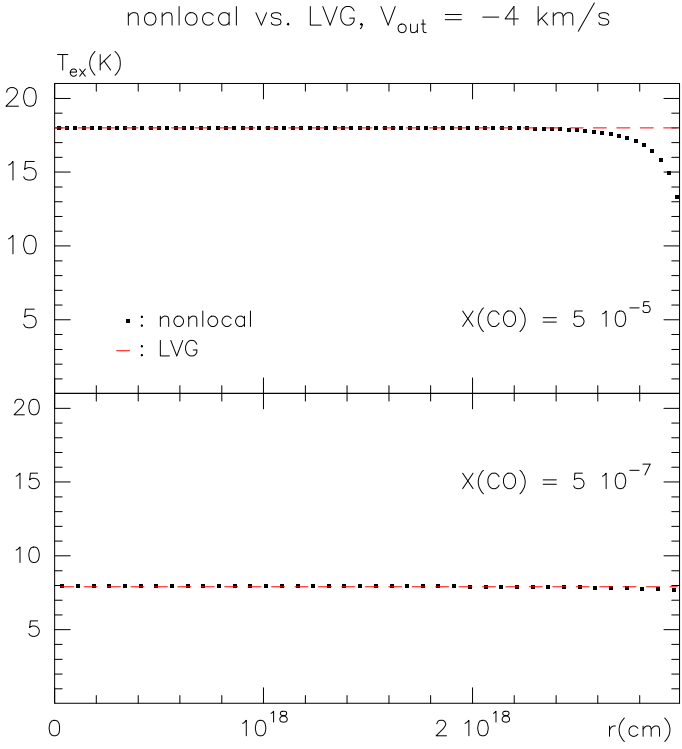}
}}
   \caption{Comparison between calculations performed using our
     non-local treatment of radiative transfer (squares) and the $LVG$
     approximation, similar to those of Fig.\ A.6, except for a higher
     value of the outer collapse velocity, in this case equal to -4
     \kms. The very low opacity case is identical to that of Fig.\ A.6
     (lowest panel) and is not displayed again.}
              \label{}%
    \end{figure}

   \begin{figure}
   \centering \rotatebox{0}{\resizebox{9cm}{!}{ 
\includegraphics{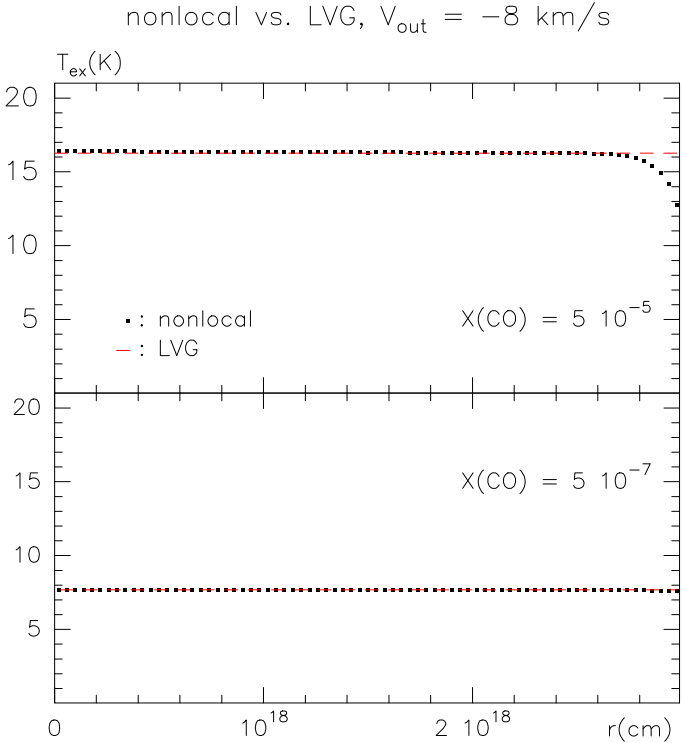}
}}
   \caption{Same as Fig.\ A.7 but for an outer collapse velocity of
     -8 \kms.}
              \label{}%
    \end{figure}

We performed calculations in other cases. In Figs.\ A.7 and A.8, we can
see the same model but assuming a higher outer collapse velocity
$V_{\rm out}$ of -4 and -8 \kms. As expected, the $LVG$ predictions are
still better. Again, for lower opacities the coincidence is absolute.

   \begin{figure}
   \centering \rotatebox{0}{\resizebox{9cm}{!}{ 
\includegraphics{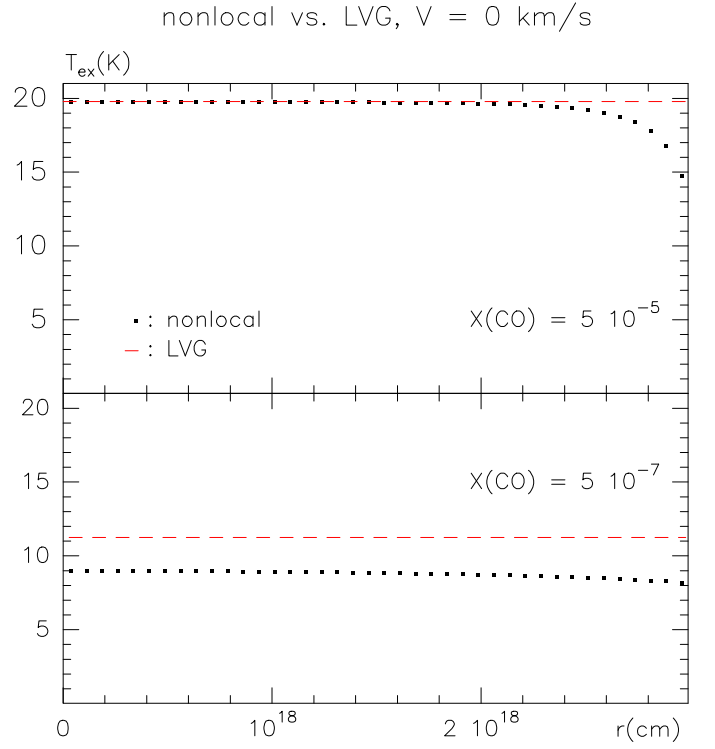}
}}
   \caption{Same as Fig.\ A.7 but without the macroscopic velocity field.}
              \label{}%
    \end{figure}

We also repeated the calculations for the case where the macroscopic
velocity is equal to zero, Fig.\ A.9, and the $LVG$ approach is forced
to maximum. (As said before, in this case the local velocity dispersion
in fact substitutes the macroscopic velocity in our $LVG$
calculations.) For high optical depths (high abundance) the agreement
is good, because the line is practically thermalized (and, of course,
the agreement is very good for very low opacities). But for cases with
moderate opacities, where the radiative transfer effects are relevant,
there is a significant difference between the non-local and $LVG$
codes; in any case, the difference is not extreme, only of about 20\%
in terms of $T_{\rm ex}$(2$-$1), a value that could be improved if we
changed the definition of the $LVG$-equivalent velocity.

   \begin{figure}
   \centering \rotatebox{0}{\resizebox{9cm}{!}{ 
\includegraphics{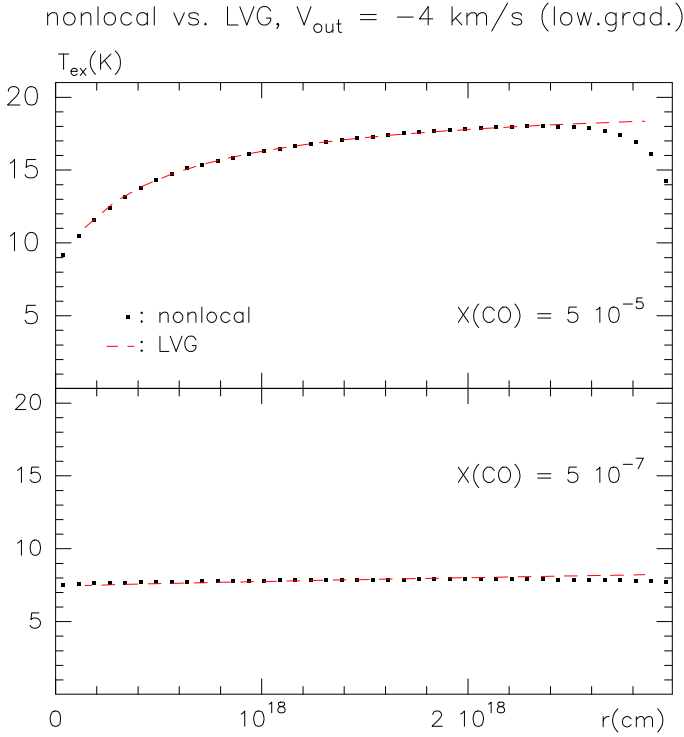}
}}
   \caption{Same as Fig.\ 7 but assuming a slowly increasing
     macroscopic velocity, defined by a constant logarithmic gradient
     of the velocity of 0.1. 
}
              \label{}%
    \end{figure}

We also performed a calculation similar to our standard case (Bernes
nebula model), but with a slowly increasing value of the radial
velocity (Fig.\ A.10). This is particularly interesting for studying
AGB circumstellar envelopes, which present a velocity law of this kind.
We considered a law $V(r)$ = $V_{\rm out} \times (r/R_{\rm
  out})^\epsilon$. A constant logarithmic velocity gradient was chosen
because this is the parameter that actually enters the calculation of
the $LVG$ escape probability. We took a quite extreme case with
$\epsilon$ = 0.1. In a law of this kind, the variation of the velocity
along the radial direction is slow and all spherical layers interact
radiatively in that direction (not in the tangential one). Results in
Fig.\ A.10 show a good behavior of the $LVG$ approach, except for the
very outer regions. Here $r/V$ is not constant across the nebula.

Finally, we performed calculations for higher-$J$ transitions with our
non-local code and compered them with $LVG$ results. We show here as an
example calculations of the \jsc\ excitation for our standard cloud
model (from Bernes 1979), but for variable CO abundances and with ten
rotational levels. (We verified that the results for \jdu\ in these
calculations are almost identical to those previously reported with
only six rotational levels.) We show the results in Fig.\ A.11. Except
for the edge effects in the very optically thick case, the $LVG$
results are identical to those from the non-local treatment. The
excitation temperature increases close to the cloud edge in the
non-local calculations for high abundance, which is because transitions
below \jsc\ are very opaque and those above it are optically thin and,
therefore, photon cascades tend to relatively overpopulate levels with
$J$ $\sim$ 6 in the outer layers.  This test is very demanding, because
the opacity is very high for $J$=5--4 and lower transitions but the
excitation state is low for high-$J$ levels (the energy of the $J$=6
level is 116 K but the kinetic temperature in the model is just 20
K). The comparable excitation given by the non-local and $LVG$
calculations confirms the accuracy of our calculations also for
high-$J$ levels and even for strong underexcitation. Similar tests are
difficult to perform by comparison with previous works because of the
lack calculations for these high transitions.

   \begin{figure}
   \centering \rotatebox{0}{\resizebox{9cm}{!}{ 
\includegraphics{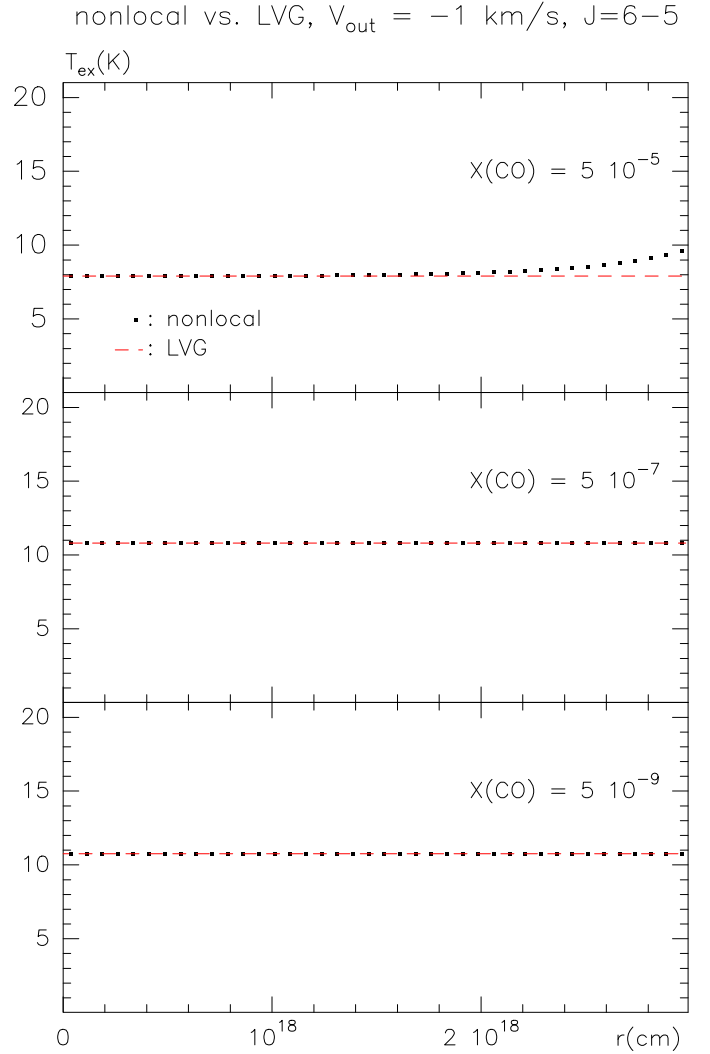}
}}
   \caption{Same as Fig.\ A.6 but for transition \jsc\ and including a
     total of ten rotational levels.  }
              \label{}%
    \end{figure}

In summary, we have seen that the $LVG$ approximation is quite accurate
in producing the excitation conditions of molecular gas 
for clouds showing radial velocities, even if they are very low and the
conditions of the approximations are in fact not satisfied. The good 
agreement of our calculations with the $LVG$ ones in all reasonable
cases must also be considered as an independent test of our non-local
code.

\subsection{Axisymmetric nebulae}

We applied our 2-D code to clouds with axial, but without spherical
symmetry. In all cases discussed here, we assumed symmetry with respect
to the equatorial plane. We considered as basic example the same
physical conditions as in previous subsections, i.e., the Bernes (1979)
nebula model with constant density, temperature, and CO abundance [$n$
  = 2 10$^{-3}$ cm$^{-3}$, $T$ = 20 K, $X$(CO) = 5 10$^{-5}$], and
radial velocity increasing linearly up to a value of 1 \kms\ in the
nebula limit. One of the radii (equatorial or axial) is kept equal to 3
10$^{18}$ cm and the other is decreased to 10$^{18}$ cm. The result is
a strongly prolate or oblate structure, in which the velocity gradients
are different in the equatorial and axial directions (while the final
velocities are the same).

We show the resulting $T_{\rm ex}$(2--1) in Figs.\ A.12 and A.13 (red
squares), continuous red lines join points along the equator and along
the axis. For comparison we show the results from the spherical cases
with cloud radii of 10$^{18}$ and 3 10$^{18}$ cm (continuous black
lines); as expected, the prolate and oblate structures show
intermediate excitations between these two extreme cases. The points
with the lowest values of $T_{\rm ex}$ are always close to the edge.

   \begin{figure}
   \centering \rotatebox{0}{\resizebox{9cm}{!}{ 
\includegraphics{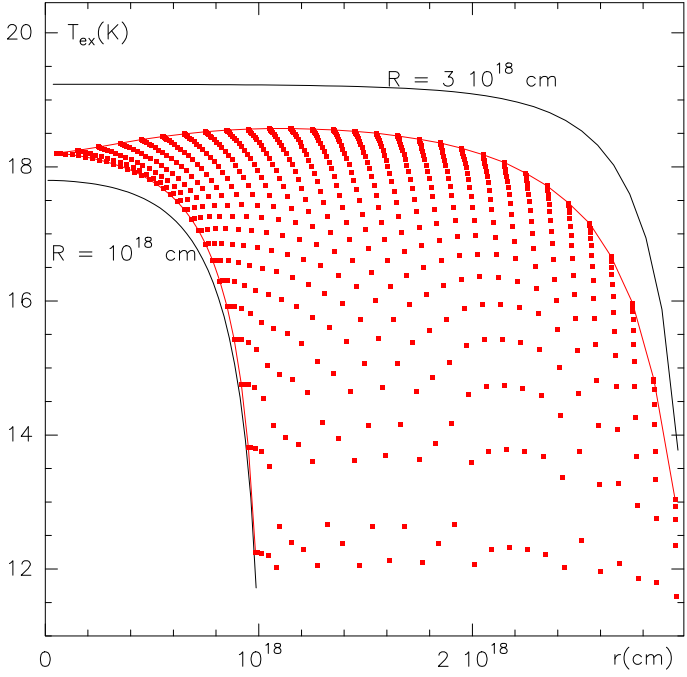}
}}
   \caption{Red squares: 2-D calculations of the \jdu\ excitation
     temperature performed for a model cloud similar to that discussed
     previously (the Bernes model cloud) but showing a strongly prolate
     structure. The red lines join points along the axis (those
     reaching higher values of the distance to the center, $r$) and
     along the equator.  Continuous black lines: calculations for
     spherical clouds with radii equal to the maximum and minimum
     radius of the prolate cloud.}
              \label{}%
    \end{figure}

   \begin{figure}
   \centering \rotatebox{0}{\resizebox{9cm}{!}{ 
\includegraphics{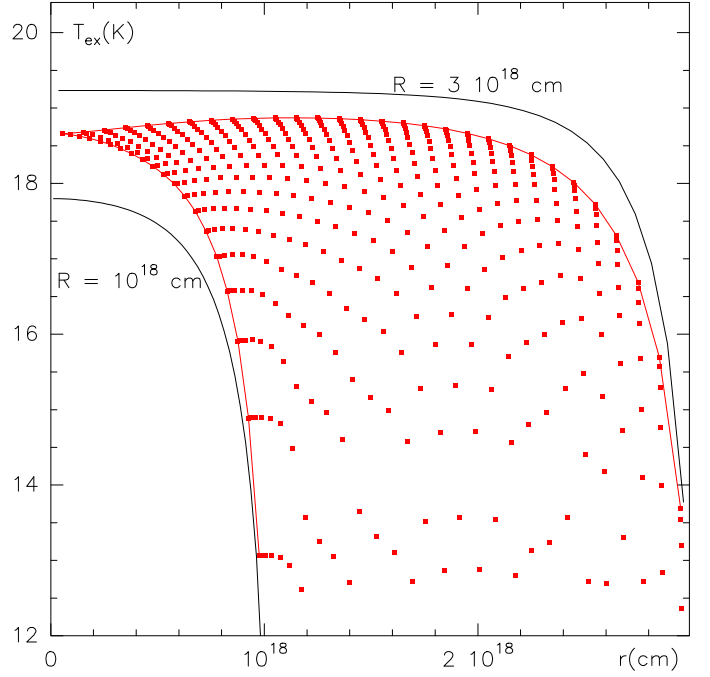}
}}
   \caption{Red squares: 2-D calculations of the \jdu\ excitation
     temperature performed for a model similar to that of Fig.\ A.12 but
     now showing a strongly oblate structure. The red lines join points
     along the equator (those reaching higher values of the distance to
     the center, $r$) and along the axis.  Continuous black lines:
     calculations for spherical clouds with radii equal to the maximum
     and minimum radius of the prolate cloud.}
              \label{}%
    \end{figure}

Finally, we incorporated Keplerian rotation in our code (which is our
goal in order to apply it to rotating disks). We considered rotation
around the symmetry axis in our oblate structure (as defined above) and
the rotation velocity was assumed to vary inversely with the
square-root of the distance to the axis, with a velocity of 0.3 \kms in
the outer regions, and reaching values higher than 4 \kms\ in the
innermost code cells. (This field would correspond to rotation around a
central compact mass of a few solar masses, but does not really
represent a stable Keplerian movement, because the total mass of the
cloud is already larger than this value.) The results can be seen in
Fig.\ A.14; again we represent the $J$=2--1 excitation temperatures
derived for the rotating cloud and for our standard spherical clouds
(continuous black lines), as in Figs.\ A.11 and A.12.

We recall that to facilitate the comparison with previous figures, we
have assumed constant density. That is usually not the case in actual
rotating disks; when we assume the density to increase toward the
center (with 1/$r$ or 1/$r^2$), the excitation is high in the central
regions and the $J$=2--1 line thermalizes in them.

We also represent in Fig.\ A.14 predictions from our $LVG$ code (dashed
red line). For these calculations, we took the absolute values of the
distance to the center and of the tangential velocity to calculate the
factor $r/V$ (as discussed in Sects.\ A.3, including the correction to
the velocity due to the local dispersion) and $\epsilon$ =
1. Evidently, the $LVG$ results give a rough idea of the expected
excitation temperatures, but errors can be very large in this case
(larger than 30\%), even if we just compare the $LVG$ results with the
non-local predictions for points along the equator (squares joined by
the upper red line).  The reason is that in rotating clouds radiative
interaction between very distant points can take place, in a complex
long-distance interaction pattern, and the $LVG$ approximation is not
valid at all. Particularly in inner points, where rotation is fast, the
$LVG$ procedure implies that radiative interaction only takes place
within a very small region, though in reality even line radiation from
the edges of the cloud can be absorbed in these innermost layers.

In summary, the $LVG$ calculations have been found to give a good
representation of the molecular excitation in spherical and
axisymmetric clouds showing radial velocity fields, even in extreme
conditions (including the case without a macroscpic velocity
field). However, in a rotating cloud the $LVG$ results only give an
idea of the true $T_{\rm ex}$ values, particularly in the inner regions
that are in fast rotation. It is difficult to imagine how the standard
$LVG$ formalism could be modified to obtain a better approximation to
the radiative transfer in this case.

   \begin{figure}
   \centering \rotatebox{0}{\resizebox{9cm}{!}{ 
\includegraphics{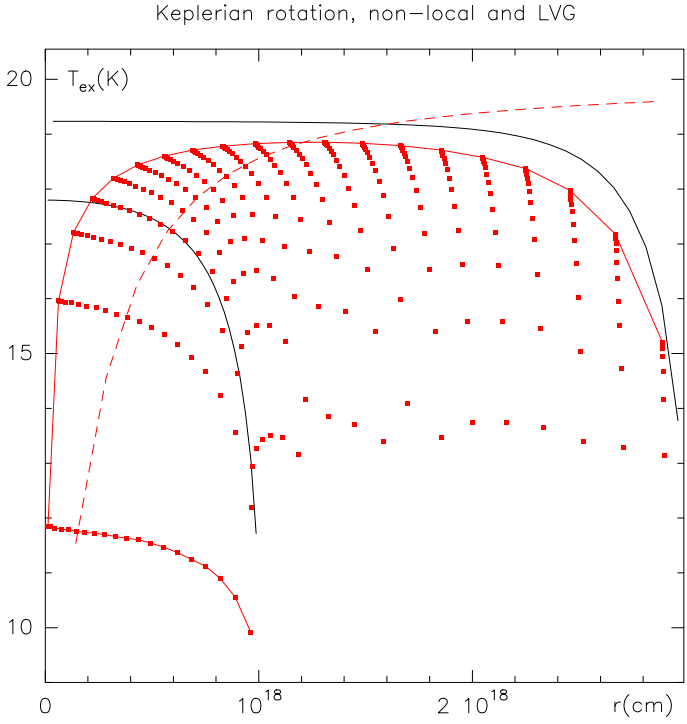}
}}
   \caption{Red squares: Calculations similar to those shown in
     Fig.\ A.13 but assuming that the oblate structure is in Keplerian
     rotation. The dashed red line represents $LVG$ calculations for
     the distance and absolute value of the velocity used here, which do
     not approximate the 'exact' code results.}
              \label{}%
\end{figure}

\section{Line predictions for alternative models of the Red Rectangle}

We investigated the variations of the predicted intensity of the lines
discussed here as a function of the physical conditions in the disk,
notably of the gas temperature, which is the parameter discussed in
more detail in this paper. In Sect.\ 4.1, we have already presented the
line intensity variation if we include a central region (at a distance
to the axis smaller than 2 10$^{15}$ cm) with temperature and density
higher than in our standard model by a factor 2. Fig.\ 2 showed that
this new component can help to explain the high line-wing emission in
intermediate-$J$ transitions, but yields too strong emission in the
\jdsq\ line, therefore the problem of the too high line-wing intensity
persists. We present here a similar case, in which the values of
$T(R_{\rm kep})$ are the same as in our standard model, but we increase
the slope of the dependence of $T$ with the distance to the center from
1 to 1.3. This is equivalent to an increase in the temperature by 20\%
at the point $R$ = $R_{\rm kep}$/2 and by 50\% at 2 10$^{15}$ cm from
the star. Results are shown in Fig.\ B.1. Clearly, the predicted
intensity for the \jdsq\ transitions is already too high and this new
law leads to a significant increase in the \jsc\ line wings. As
discussed in Sect.\ 4.1, models of rotating disks of this kind cannot
explain the \jsc\ line wings and the \jdsq\ profile.

   \begin{figure}
   \centering \rotatebox{0}{\resizebox{9cm}{!}{ 
\includegraphics{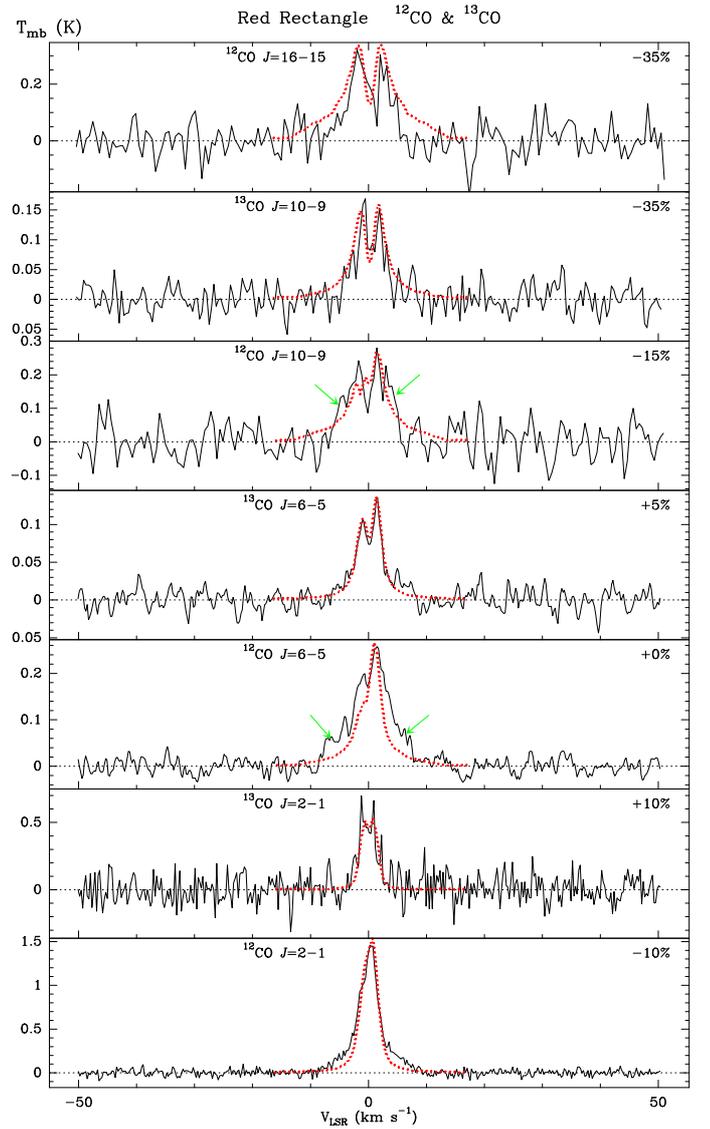}
}}
   \caption{Profiles of the observed mm- and submm-wave transitions
     (black) and the predictions of our code (red points), assuming
     that the slope of the temperature profiles increases to 1.3
     instead of 1, as in our standard model, Fig.\ 1, Table 2. The
     symbols have the same meaning as in Fig.\ 1; note the values of
     the multiplicative parameter.}
              \label{}%
    \end{figure}

We investigated the deviations of the predictions from the observed
profiles assuming changes in the temperature from our best-fitting
model. In Figs.\ B.2 and B.3, we show predictions assuming that the
temperature varies by $\pm$ 20\% with respect to our standard model,
while the rest of the parameters remain the same. These variations are
sufficient to yield predictions incompatible with the data (though not
by a very large factor), taking the calibration uncertainties into
account, and can be considered in these conditions as a measure of the
uncertainty in the estimated temperature values.

   \begin{figure}
   \centering \rotatebox{0}{\resizebox{9cm}{!}{ 
\includegraphics{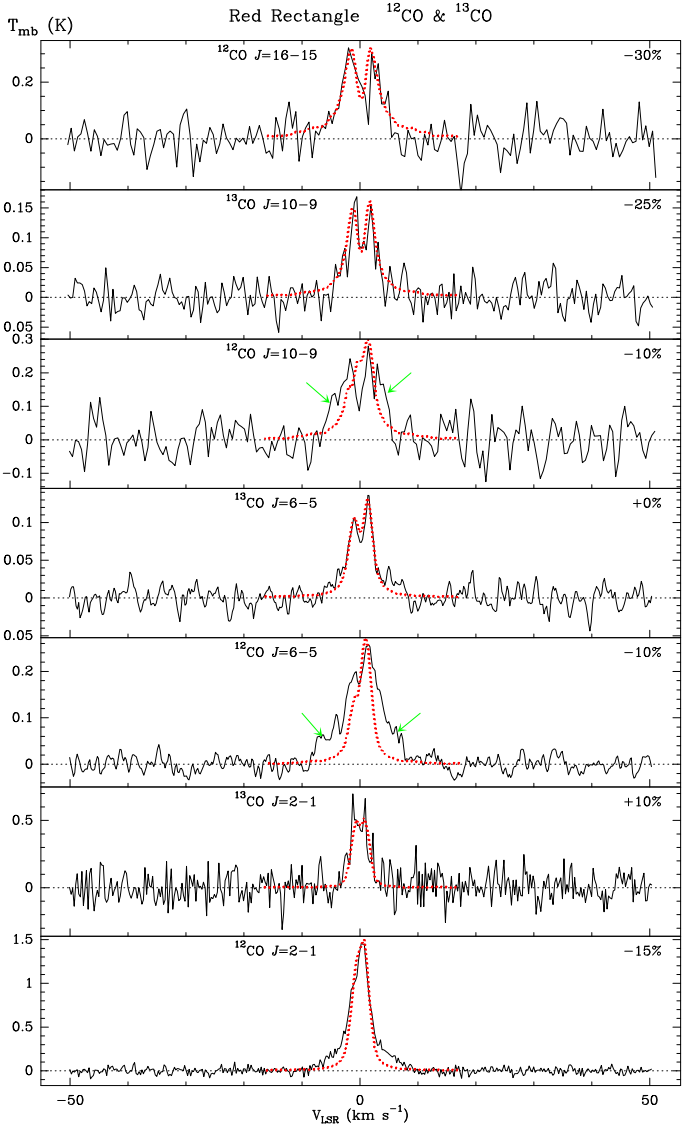}
}}
   \caption{Profiles of the observed mm- and submm-wave transitions
     (black) and the predictions of our code (red points), assuming
     that the temperature increases by 20\% with respect to our
     standard model, Fig.\ 1, Table 2. The symbols have the same
     meaning as in Fig.\ 1; note the values of
     the multiplicative parameter.}
              \label{}%
    \end{figure}

   \begin{figure}
   \centering \rotatebox{0}{\resizebox{9cm}{!}{ 
\includegraphics{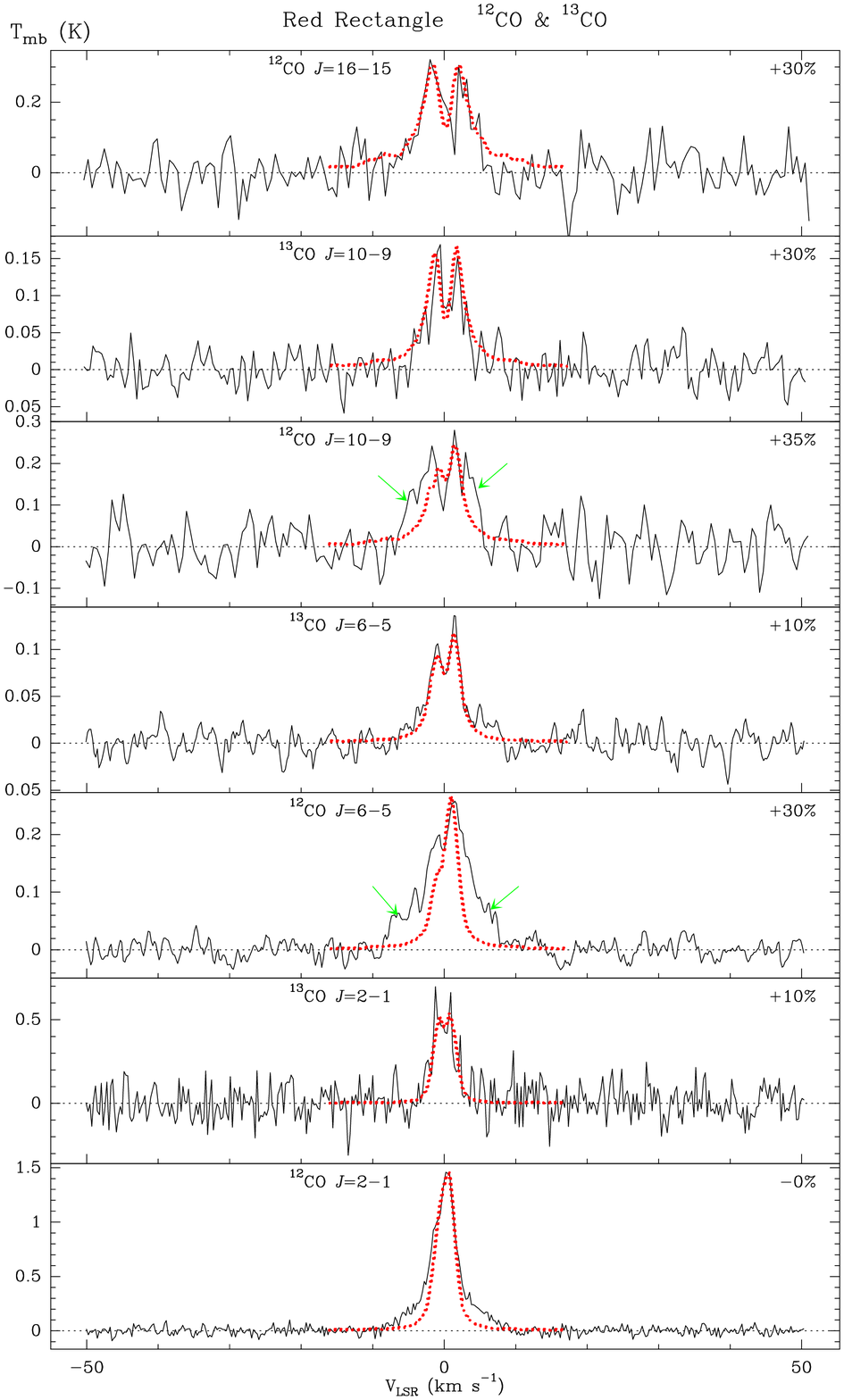}
}}
   \caption{Profiles of the observed mm- and submm-wave transitions
     (black) and the predictions of our code (red points), assuming
     that the temperature decreases by 20\% with respect to our
     standard model, Fig.\ 1, Table 2. The symbols have the same
     meaning as in Fig.\ 1; note the values of
     the multiplicative parameter.}
              \label{}%
    \end{figure}

The low-$J$ transitions depend very little on the kinetic temperature
for the investigated cases.

Finally, we considered a nebula model in which both the density and
temperature vary. We assumed that the density increases by a factor 2
and at the same time that the temperatures decreases by a 30\%. The
results are very similar to those shown in Fig.\ B.3, the comparison
with the observations is not acceptable, but now for a variation in the
temperature of a 30\%. We conclude that allowing variations in the
density relaxes the uncertainties in determining the temperature, but
not by a large factor, from 20\% to 30\% in this case. Deeper changes
in the model would contradict with our initial intention of keeping the
nebula model deduced from mm-wave maps (Bujarrabal et al.\ 2005)
invariable as far as possible .

\end{document}